\documentclass[twocolumn, pra, showpacs, superscriptaddress, floatfix, longbibliography, 10pt]{revtex4-2}
\usepackage[english]{babel}

\usepackage[T1]{fontenc}
\usepackage{lmodern}
\usepackage{amsmath}
\usepackage{amssymb}
\usepackage{graphicx}
\usepackage{xcolor}
\usepackage{placeins} 
\usepackage{booktabs}
\usepackage{lipsum}

\usepackage[colorlinks=true]{hyperref} 
\hypersetup{
    colorlinks=true,
    linkcolor=blue,
    citecolor=blue,
    urlcolor=blue,
    bookmarksdepth=0
}

\newif\ifshowmb
\showmbtrue      

\newcommand{\score}{\ensuremath{\phi}}

\begin{document}
\title{Modeling Protein Evolution with Generative Models:\\ from Extant Sequence Data to Evolutionary Dynamics}

\author{Matteo Bisardi}
\thanks{These authors contributed equally to this work.}
\affiliation{Michael Smith Laboratories, University of British Columbia, Vancouver, British Columbia, Canada}

\author{Leonardo Di Bari}
\thanks{These authors contributed equally to this work.}
\affiliation{DISAT, Politecnico di Torino, Corso Duca degli Abruzzi, 24, I-10129, Torino, Italy}
\affiliation{Sorbonne Universit\'e, CNRS, Computational, Quantitative and Synthetic Biology, 75005 Paris, France}

\author{Saverio Rossi}
\thanks{These authors contributed equally to this work.}
\affiliation{Dipartimento di Fisica, Sapienza Universit\`a di Roma, Piazzale Aldo Moro 5, 00185 Rome, Italy}

\author{Karol Buda}
\affiliation{Michael Smith Laboratories, University of British Columbia, Vancouver, British Columbia, Canada}

\author{Francesco Zamponi}
\email{Email: francesco.zamponi@uniroma1.it}
\affiliation{Dipartimento di Fisica, Sapienza Universit\`a di Roma, Piazzale Aldo Moro 5, 00185 Rome, Italy}

\author{Martin Weigt}%
\email{Email: martin.weigt@sorbonne-universite.fr}
\affiliation{Sorbonne Universit\'e, CNRS, Computational, Quantitative and Synthetic Biology, 75005 Paris, France}
\affiliation{Institut Universitaire de France (IUF)}

\date{\today}

\begin{abstract}
Protein sequences carry a record of evolutionary history shaped by mutation, selection, drift, and epistasis. Recent generative models trained on homologous sequence families offer a new way to read this record: they define probabilistic landscapes that score sequences, generate viable variants, and capture constraints that are difficult to measure experimentally. In this review, we discuss how such landscapes can be used not only for protein design or mutation-effect prediction, but also for modeling evolutionary dynamics. We focus particularly on Direct Coupling Analysis as an interpretable and experimentally validated framework, while placing it in the broader context of generative sequence modeling. We first describe how generative sequence landscapes are inferred and assessed, then review how they can be coupled to population-genetic or substitution-model dynamics to simulate protein evolution across experimental and phylogenetic timescales. Applications include viral evolution, laboratory drift experiments, historical contingency, entrenchment, epistatic drift over time, and long-term sequence-space exploration. We conclude by discussing open challenges, including score–fitness calibration, phylogenetic structure, codon-level mutation biases, indels, and the integration of experimental data.
\end{abstract}

\maketitle
\onecolumngrid
\begingroup
\makeatletter
\let\l@subsubsection\@gobbletwo
\makeatother
\tableofcontents
\endgroup
\clearpage
\twocolumngrid


\section{Introduction}

Throughout billions of years of evolutionary history, organisms have continuously adapted to ever-changing ecological niches. 
During this time, natural selection has generated an astounding diversity of molecular solutions to the challenges and opportunities encountered in the environment.
Consequently, every extant organism carries a record of its evolutionary history written at the molecular scale. 
Thus, by studying the underlying variation in these biomolecules, it is possible to gain insight into the fundamental principles governing evolutionary dynamics.

While the physical sciences have historically relied on universal mathematical laws to explain the natural world, our understanding of biology has often been distorted by system-specific complexity and apparent stochasticity stemming from historical contingency that arises during evolution. This distinction has frequently made it difficult to apply the rigorous, first-principles approach of physics to biological systems. 
Nevertheless, physicists' and mathematicians' interest in biological processes dates back many decades, leading to the application of mathematical methods across multiple biological disciplines.

For a long time, however, these approaches remained largely theoretical. A lack of extensive, reliable genetic data limited the ability to directly confront mathematical models with reality, let alone with experiments. However, this situation has changed dramatically over the past few decades, with an explosion in both the quantity and quality of molecular biological data. In particular, the availability of large-scale protein sequence datasets, stemming from improvements in next-generation sequencing throughput and costs~\cite{van2014ten}, now enables a direct, high-resolution comparison between models and the actual evolutionary history of amino acid sequences.

These developments in sequencing technology and data availability have paralleled the rise of powerful generative architectures in machine learning, inspired by physics like Boltzmann machines~\cite{ackley1985learning} and, more recently, rooted in natural language processing like transformers~\cite{vaswani2017attention}. Just as unsupervised generative models trained on vast textual corpora have demonstrated remarkable capabilities in generating coherent text, similar advanced statistical tools are now being applied to the ``language'' of proteins.

In this biological context, generative sequence models are trained on large sets of homologous protein sequences that display high sequence variability while retaining similar biological functions. These models have been applied to a wide range of problems, including protein structure prediction~\cite{lin2023evolutionary}, cellular localization~\cite{stark2021light}, mutation-effect prediction~\cite{cheng2024zero}, and the extraction of sequence representations useful for protein engineering~\cite{madani_large_2023}. By contrast, comparatively less attention has been devoted to the use of these models for studying and simulating protein evolution itself, even though evolution is precisely the process that generated these sequences and the underlying variation on which these models are trained. In this review, we place this question at the center. In other words, we move the focus from static properties and applications of generative sequence models to dynamical ones, in which sequence landscapes can be navigated \textit{in silico}.

More specifically, we argue that the synthetic fitness landscapes defined by generative sequence models provide a suitable effective space in which to model protein evolution. Unlike biophysical models that require the explicit encoding of structural or biochemical constraints, generative models infer the underlying constraints of protein ``fitness'' directly from the statistical patterns found in nature. As such, they represent effective and simplified descriptions of the complex genotype-to-fitness relationship that characterizes proteins.

In this review, we introduce foundational concepts and methods of this emerging field and highlight recent advances arising from the synthesis of data-driven sequence landscapes and classical protein evolution modeling. In particular, we discuss how to integrate generative sequence landscapes with classical population genetic and substitution models to produce realistic simulations of protein evolution across a range of temporal and phylogenetic scales.
We refer to this general methodology as \textit{generative protein evolution}. 

While the methods and results discussed in this review apply in principle to every class of generative models, particular emphasis is placed on models based on Direct Coupling Analysis (DCA), which have thus far provided some of the most quantitative and interpretable applications of generative models to understand the evolutionary dynamics of protein sequences.

\subsection*{Review structure}

In \autoref{chap:genseq}, we introduce generative sequence landscapes and how to construct them from collections of homologous protein sequences. 
In particular, we focus on a specific model, direct coupling analysis, in its generative version. 
We argue that it is a good proxy for realistic fitness landscapes, and we summarize a body of results that showcase its predictive ability and relevance to modeling evolution.
In particular, we show that DCA captures epistatic constraints from sequence data, and we discuss their relevance for experimental fitness and natural mutability prediction as well as artificial protein generation.

In \autoref{chap:evodin}, we present generative protein evolution models. 
We discuss two main approaches to modeling evolution on fitness landscapes: population genetics and substitution models.
We outline their key differences and highlight how generative models can be incorporated into these evolutionary frameworks. 
We further establish a connection between classical substitution models and the Markov Chain Monte Carlo (MCMC) dynamics commonly used in statistical physics. 
This allows us to define, more generally, a class of substitution models that we call Generative Substitution Models (GSM).
This methodological section aims to bridge these complementary perspectives and to develop a common language.

In \autoref{chap:res}, we collect and summarize the results published in the field of GSMs and generative population genetics modeling. We identify a few common themes and modeling goals, and we separate results based on the availability of experimental data to substantiate the predictions.
The most timely application of these models relies on simulations and the prediction of short-time evolutionary dynamics, with a focus on pathogenic proteins. 
The comparison of synthetic results with natural and experimental observations shows strong correlations.
A more speculative and theoretical stream of works instead makes use of models to move beyond the time scales accessible by experiments and give insights into the long-time features of protein evolution.

In \autoref{chap:persp}, we discuss future directions in the field.
On one hand, we argue about what can be done in order to improve the performance of the generative modeling (e.g., via integration of experimental results).
A deeper interconnection between experiments and modeling can benefit both sides of the problem, with models becoming more precise and experiments more efficient.
We then delve into some novel applications that benefit from the use of generative sequence landscapes to reproduce protein evolution.

\section{Generative sequence landscapes}
\label{chap:genseq}

\begin{figure*}
    \centering
    \includegraphics[width=0.95\textwidth]{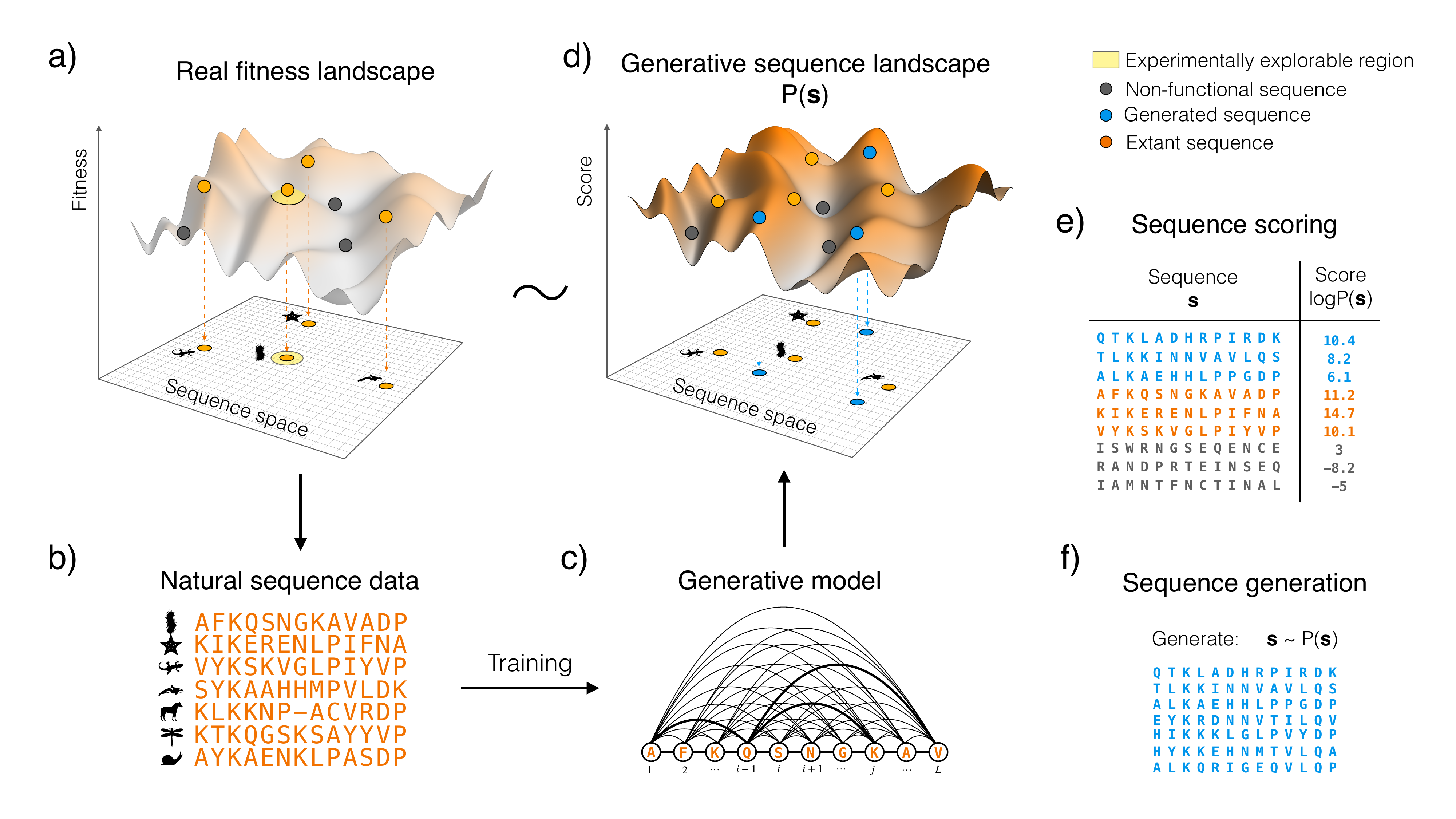}
    \caption{ \textbf{Schematic representation of generative sequence landscape.} 
    (a) Schematic representation of a protein fitness landscape and its corresponding sequence space. Each point on the fitness landscape represents a protein sequence, and its height represents fitness. Extant natural sequences occupy functional regions of the landscape, whereas non-functional sequences lie in low-fitness regions. Experimental approaches can typically probe only restricted local regions of sequence space around known functional sequences, as depicted by the yellow shaded region. On the contrary, the fitness landscape as a whole is not observable, as signaled by its opacity.
    (b) Extant natural homologous sequences can be collected and organized in protein families that are used as training data for generative models.
    (c-d) After training, the model defines a probability distribution $P(\mathbf{s})$ over sequence space, represented here as a reconstructed sequence landscape. High-probability sequences correspond to regions compatible with the statistical constraints of the protein family. Extant sequences (orange) occupy high probability regions, but the model can also interpolate between them (light blue).
    (e) Generative sequence models can be used to score arbitrary sequences through the log-probability $\log P(\mathbf{s})$. Designed and natural sequences have high scores, non-functional sequences have low scores.
    (f) The inferred sequence landscape can be used to generate new sequences by sampling $\mathbf{s} \sim P(\mathbf{s})$.
    }
    \label{fig:fitnessland}
\end{figure*}

\subsection*{Motivation}
Fitness landscapes are simplified, conceptual representations that connect genotypes across a multi-dimensional sequence space to fitness (or phenotype), describing how mutation, selection, and genetic drift shape evolving proteins~\cite{fragata2019evolution}. For much of its history, the study of fitness landscapes was constrained by a lack of breadth in empirical fitness data, forcing a reliance on idealized theoretical models~\cite{kingman1978simple, kauffman1987towards, kauffman1989nk, fisher1999genetical, tenaillon2014utility}, which have, nonetheless, successfully explained several general features of protein evolution~\cite{pahujani2025complexity, srivastava2026evolution}. However, the lack of empirical validation for such models has historically limited their application to specific biological systems~\cite{fragata2019evolution, ribeca2026simplesignepistasisevolutionary}.

While Deep Mutational Scanning (DMS) experiments have begun to bridge this gap by characterizing the local landscape of proteins through the assaying of thousands of single mutational variants in parallel~\cite{fowler2014deep, firnberg2014comprehensive}, these measurements are inherently constrained to the immediate neighborhood of a fixed reference sequence. Even combinatorial experiments that test multiple amino acid combinations on a specific reference protein remain restricted to narrow, local regions~\cite{poelwijk2019learning, papkou2023rugged, buda2023pervasive} and their exploratory scale is limited by the exponential increase in sequence space proportional to the protein's length.
As a consequence, it is very hard to capture experimentally the effects of epistasis, \textit{i.e.}, the phenomenon in which a mutation's impact depends on its genetic context, in shaping evolution over long timescales ~\cite{de2014empirical, starr2016epistasis, johnson2023epistasis, miton2016mutational}. 

While an exhaustive experimental characterization of the entire fitness landscape of a gene is both unfeasible and unnecessary—as natural selection confines functional proteins to a tiny fraction of the total sequence space—another approach is possible.
To efficiently explore this restricted landscape, unsupervised data-driven statistical models—such as DCA—have emerged as a powerful alternative to purely experimental approaches. By leveraging the exponential growth of unlabeled homologous sequences deposited in public databases~\cite{blum2024interpro}, these generative models can infer probability distributions over the biologically relevant regions of sequence space, providing quantitative descriptions of the evolutionary constraints acting on protein families. Throughout this review, we refer to these inferred probabilistic representations of fitness landscapes as Generative Sequence Landscapes.

The logic behind this construction is summarized in Fig.~\ref{fig:fitnessland}. 
The complete fitness landscape of a protein family cannot be empirically mapped, and experimentally accessible regions constitute only a minuscule fraction of sequence space (Fig.~\ref{fig:fitnessland}a). Natural homologous sequences have survived selection and therefore occupy high-fitness regions of the landscape. 
As such, they provide informative, albeit evolutionarily biased, samples of its functional regions (Fig.~\ref{fig:fitnessland}b). Generative models are trained on these sequences to learn statistical constraints across sites (Fig.~\ref{fig:fitnessland}c), defining an explicit probability distribution \(P(\mathbf{s})\) over sequences. 
This bypasses explicit mechanistic modeling of biochemical and biophysical processes~\cite{liberles2012interface, sikosek2014biophysics, bastolla2017evolution, echave2017biophysical}. 
The inferred distribution constitutes a generative sequence landscape, approximating the functional regions of the underlying fitness landscape (Fig.~\ref{fig:fitnessland}d).
Although Fig.~\ref{fig:fitnessland} provides an intuitive visualization of a fitness landscape, it is important to emphasize that it is only a schematic representation. Because sequence space is highly dimensional, the true fitness landscape cannot be visualized directly and is substantially more complex than the low-dimensional surface depicted here.


While early formulations of such models were rooted in statistical physics~\cite{lapedes1999correlated, weigt2009identification}, the field has expanded to include a wide range of deep learning architectures capable of modeling either specific protein families or the broader protein universe~\cite{bjerregaard2025foundation}.


From the probability \(P(\mathbf{s})\), the quality of a sequence $\mathbf{s}$ can be defined as its log-likelihood, which we refer to as the score  $\score(\mathbf{s})$:
\begin{equation}\score(\mathbf{s}) = \log P(\mathbf{s}).\label{eq:1}\end{equation}
In this framework, high-probability regions—and thus high scores—are assumed to correspond to high biological fitness. We note that, in the context of DCA, the score is often defined as the negative log-likelihood and therefore corresponds to an energy in the physics sense.

This inferred distribution allows for two primary applications: first, scoring arbitrary sequences via the log-probability $\score(\mathbf{s})$ (Fig.~\ref{fig:fitnessland}e) and second, sampling new variants ($\mathbf{s} \sim P(\mathbf{s})$) to generate novel functional sequences absent from the training set (Fig.~\ref{fig:fitnessland}f).
Within this framework, designed and natural sequences are expected to yield high scores, whereas non-functional sequences occupy low-probability regions of the inferred landscape.

Broadly, a \textit{generative} model is expected to produce novel data statistically indistinguishable from the training set. In protein contexts, this is typically benchmarked by comparing the statistical properties of sampled sequences to natural datasets~\cite{figliuzzi2018pairwise, mcgee2021generative}. 
However, while statistical similarity is a useful proxy, the definitive measure of a model’s performance is its ability to produce protein variants that are functionally viable in vitro or in vivo while suggesting sequences that are divergent from those used in the training data. The relationship between model scores and actual fitness is rarely monotonic. Specifically, in rapidly mutating viruses subject to diverse immune pressures, sequence landscapes inferred from patient databases have been shown to correctly reflect the rank order of replicative fitness of mutant strains~\cite{shekhar2013spin}. While fitness can be linearly translated into a score in such specific cases, a nonlinear relationship often persists in more generic biological settings~\cite{kaltenbach2014dynamics}.

\subsection*{From sequence data to sequence landscapes}
In this section, we provide an overview of the methodologies and approaches employed to approximate protein fitness landscapes with generative models. We focus in particular on DCA, which will be the main model discussed in our review. For in-depth discussions about the model and its applications, we refer the readers to Refs. ~\cite{morcos2014direct, levy2017potts, cocco2018inverse}.

There are several reasons for using DCA to model protein evolution. Firstly, this model has a very simple and interpretable architecture. It is also trained exclusively on aligned sequences, which are central to evolutionary biology and phylogenetics.
Moreover, since it was first applied to protein sequence data almost twenty years ago \cite{weigt2009identification}, DCA has been extensively validated and successfully applied across a wide range of biological problems. 
As a result, it has become the model of choice for much of the work on generative protein evolution conducted to date. 

\subsubsection*{Training data}

We begin with a brief description of the training data before introducing the models used to infer generative sequence landscapes. 
In most cases, generative models are trained on protein sequences obtained in natural organisms and grouped into protein families~\cite{blum2024interpro}. 
Although such families may include paralogs or very functionally divergent orthologs, as well as unevenly sampled sequences across evolutionary lineages and ecological niches, they are typically assumed to share a common fold and function, despite substantial sequence variability.
These sequences are then aligned into multiple sequence alignments (MSAs), which provide the input data for generative model training. The number of sequences and their variability with respect to alignment length are important to guarantee the inference of a good statistical model \cite{haldane2019influence}.

Thanks to the sequence heterogeneity of their members, sequence families contain rich information about the structural and functional constraints acting on proteins, reflected in patterns of conservation and coevolution between residues \cite{tiana2009molecular, crippa2021evolution}. 
In particular, coevolutionary correlations encode epistatic interactions arising from structural contacts and functional requirements, and have been shown to contain sufficient information to specify protein fold and function~\cite{socolich2005evolutionary}. 
By learning and reproducing these statistical patterns in an unsupervised fashion, generative models capture the constraints that define functional sequences and provide an effective representation of the underlying fitness landscape.

\subsubsection*{Direct Coupling Analysis}

One of the simplest approaches to model the statistics of aligned protein sequences is the so-called profile model, which assumes that residues in the protein are independent. 
In this framework, the probability $P(\mathbf{s})$ of a given sequence $\mathbf{s} = (s_1, \dots, s_L)$ can be factorized as the product of the probabilities at the single residues, which makes the statistical score $\score(\mathbf{s})$ the sum of the contributions of each site, namely
\begin{equation}
    \score(\mathbf{s}) = \sum_{i=1}^L h_i(s_i).
\end{equation}
A common choice for this type of models is to take the contribution of each amino acid to the total statistical score as $h_i(a) = \log(f_i(a))$, where $f_i(a)$ is the frequency with which it appears in the protein family alignment~\cite{cocco2018inverse, mcgee2021generative}.
However, this kind of architecture neglects the contribution of epistasis and thus is incapable of accounting for the context-dependence of mutational effects~\cite{bisardi2022modeling, mcgee2021generative}. 

The origin of epistasis has been explored extensively through experimental studies~\cite{lunzer2010pervasive,biswas2019epistasis,poelwijk2019learning,domingo2019causes,park2022epistatic,chen2023understanding}, and the results often reveal that direct, pairwise interactions between specific mutations are frequently too weak to be measurable individually. Rather than resulting from a few isolated strong links, epistasis typically emerges as a collective phenomenon, adding up many small contributions from multiple mutations.

Direct Coupling Analysis extends the profile model by incorporating interaction terms between any pairs of sites, hence moving beyond the assumption of site independence.
Initially designed for residue-residue contact prediction~\cite{weigt2009identification}, DCA has expanded into a robust family of methods parameterized to capture the global co-evolutionary structure of protein families~\cite{ekeberg2013improved, figliuzzi2018pairwise, trinquier2021efficient}. 
Historically, model inference schemes have progressed using mean-field approximations~\cite{morcos2011direct}, cluster expansion methods~\cite{cocco2011adaptive}, pseudolikelihood maximization (plmDCA)~\cite{ekeberg2013improved}, Boltzmann-machine learning (bmDCA)~\cite{ackley1985learning, figliuzzi2018pairwise}, and autoregressive frameworks (arDCA)~\cite{trinquier2021efficient}. 
It is important to note that only models trained on the generative objective, like bmDCA and arDCA, can be used to correctly model protein evolution. 

As for generative models in general, the score of DCA is defined through the logarithm of its probability as in Eq.\eqref{eq:1}. In its specific formulation, it is the sum of ``fields'' $h_i$ acting on single residues and ``couplings'' $J_{ij}$ acting on pairs of residues accounting for epistasis, as follows
\begin{equation}
\label{eq:dca_en}
    \score(\mathbf{s}) =  \sum_{i=1}^{L} h_i(s_i) \,\,+\sum_{1 \le i < j \le L} J_{ij}(s_i, s_j).
\end{equation}
The parameters $h_i(a)$ and $J_{ij}(a, b)$ are the ones to be inferred to fully specify the model. After inference, the number of parameters of the models is of the order of $O(q^2L^2)$. Although the number of parameters can be very high, the model is interpretable, as $J$ encodes co-evolutionary signals among protein sites and $h$ accounts for site-specific conservation.
Methods to reduce the number of parameters while retaining a comparable statistical performance have been developed as well \cite{barrat2021sparse, tsishyn2026structure}.

The most commonly employed inference method exploits Boltzmann machine learning to infer the model parameters. Thanks to efficient GPU implementations~\cite{haldane2021mi3, rosset2025adabmdca}, it is possible to construct synthetic fitness landscapes for protein sequence families with hundreds of residues. 

These kinds of models are, by construction, capable of generating batches of synthetic sequences that have the same single-site and pairwise amino acid statistics of the alignments used as a training set \cite{figliuzzi2018pairwise}. 
In addition, in the case of DCA models, when synthetic sequences were tested for a few protein families, it has been confirmed that a sizable fraction was actually functional \cite{russ2020evolution,fram2024simultaneous, fernandez2025designing,  netti2026expanding} (see below for a more detailed discussion). 
The statistical similarity to natural sequence alignments and the experimental validation certifies then that DCA models are generative in the sense that we described at the beginning of this section.

\subsubsection*{Other generative approaches}

In this section, we briefly discuss other generative sequence model architectures that can be used to construct generative sequence models and to generate novel functional proteins. These approaches can be broadly divided into models trained on aligned sequences from a single protein family and models pre-trained on large corpora of unaligned protein sequences. More detailed overviews of this rapidly evolving literature can be found elsewhere~\cite{romero2023exploring, mardikoraem2023generative, bjerregaard2025foundation}, including for semi-supervised methods that combine sequence data with experimental fitness measurements~\cite{hsu2022learning}. 

Among family-specific models, Restricted Boltzmann Machines (RBMs) are conceptually close to DCA, as they are also trained on aligned sequence data and define an explicit probability distribution over sequences. However, instead of representing epistasis through direct pairwise couplings, they introduce a layer of hidden variables that mediates interactions between residues~\cite{tubiana2019learning, shimagaki2019selection}. In this way, RBMs can capture effective higher-order dependencies while remaining relatively lightweight and simple to sample from~\cite{Decelle2025inferring, Huot2025generative, Huot2025constrained, Rehan2025design}. They can thus be viewed as an intermediate architecture between pairwise statistical models and deeper neural-network approaches.

Another important class of family-level generative models is given by Variational Auto-Encoders (VAEs). These models map sequences into a lower-dimensional latent space and learn to reconstruct them from this compressed representation. In doing so, they provide a smooth representation of sequence space in which evolutionary and functional relationships can be organized and explored~\cite{ding2019deciphering, hawkins2021generating, Ziegler2023Latent, Shukla2025Thermal}.

In parallel, recent years have seen the rapid development of deep neural-network models trained on unaligned sequences obtained from very large protein databases. These models, sometimes having billions of parameters, learn statistical regularities across the protein universe and therefore capture constraints that extend beyond any single family~\cite{rives2021biological, hesslow2022rita}. In particular, transformer-based protein language models have become highly effective tools for protein analysis and design~\cite{sevgen2025prot}. Borrowing architectures from natural language processing, they have become a central tool for mutation scoring, sequence generation, and design tasks~\cite{notin2022tranception, chandra2023transformer, wang2025comprehensive, Koehl2026deep}.

Despite their broad expressive power, these deep models are computationally much heavier and generally less convenient than DCA-like models for explicit simulations of evolutionary dynamics. To our knowledge, they have not yet been explicitly used to study protein evolution in the same way as the approaches discussed in this review, although recent deep-learning implementations have started to develop native frameworks for simulating sequence evolution along phylogenetic trees~\cite{Koehl2026deep}.

\subsection*{Biological relevance of generative sequence landscapes}

\begin{figure*}
    \centering
    \includegraphics[width=0.95\textwidth]{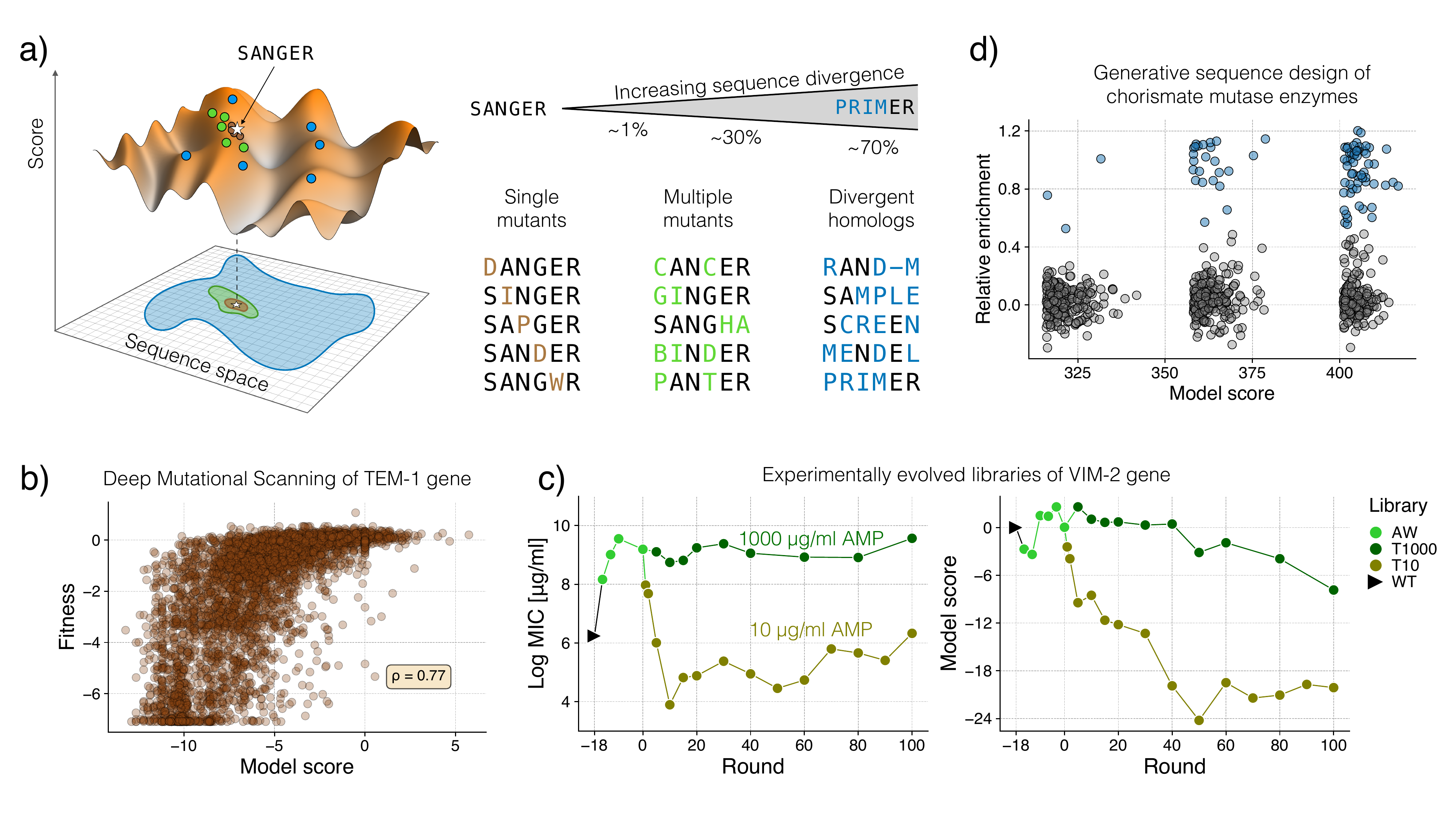}
    \caption{\textbf{
    Experimental validation of generative sequence landscapes across evolutionary scales.} 
    (a) Schematic representation of the different evolutionary scales over which sequence landscapes can be compared with and validated against experimental data. Increasingly divergent regions of sequence space with respect to the hypothetical wild-type sequence SANGER (white star) are illustrated by shaded areas of different colours.
    Word variants conceptually illustrate 
    single mutants (brown), multiple mutants within the same sequence context (green), and divergent homologs (blue).
    (b)  Predictive performance of model score differences in deep mutational scanning experiments. The panel shows a scatter plot of the change in model score relative to the wild type versus experimentally measured fitness for mutants of the TEM-1 beta-lactamase gene~\cite{firnberg2014comprehensive}. (c) Sequence landscapes can differentiate high- versus low-resistance variants of the $\beta$-lactamase VIM-2 gene across $100$ rounds of a long-term neutral genetic drift experiment \cite{erdougan2023neutral}. The left (right) panel represents the average log minimum inhibitory concentration (difference in model score) of VIM-2 variants across three different experimental conditions: adaptive walk (AW), selection at $1000 \,\mu g/ml$ of ampicillin (T1000), and selection at $10 \,\mu g/ml$ of ampicillin (T10).
    (d) Sequences generated from generative models can be functional in vivo. The panel shows the relative enrichment of chorismate mutase enzymes designed using DCA as a function of their model score~\cite{netti2026expanding}. Light-blue points correspond to sequences classified as functional, whereas dark-grey points indicate inactive sequences.
    }
    \label{fig:fig2}
\end{figure*}

Once a generative sequence landscape has been inferred, its ability to approximate real protein fitness landscapes should be assessed against experimental evidence. 
In recent years, especially for deep models, this evaluation has increasingly relied on standardized benchmark-style comparisons~\cite{notin2024proteingym}. 
Although such benchmarks provide a useful common ground for model comparison, they often reduce the problem of evaluating models to mutation-effect prediction and do not fully address the broader biological plausibility of the inferred landscape, nor the amount of epistasis that it captures. 
Here, we adopt a wider perspective, focusing on the extent to which the inferred landscapes capture features that are relevant for modeling natural protein evolution.

Two complementary strategies can be followed to test the accuracy of a model, as highlighted by Fig.~\ref{fig:fig2}(a). 
The first focuses on the local structure of the landscape, typically in the neighborhood of a reference sequence, where mutational effects can be characterized in detail. 
The second probes the landscape more globally, by testing whether the model captures broader properties of sequence space and of the family it is meant to describe. 
In the following, we argue that DCA-based sequence landscapes capture biologically relevant and evolutionarily meaningful features of protein fitness landscapes across a broad range of sequence families and biological contexts.

Before reviewing these results, it is important to clarify the meaning of the model score defined in Eq.~\eqref{eq:1}. 
Because these models are inferred from natural sequence databases, the scores reflect statistical trends rather than absolute biological fitness or pure biochemical functionality. 
Nevertheless, by comparing these inferred landscapes against experimental data and observed natural variability, we gain a robust framework for assessing model quality.

\subsubsection*{Fitness prediction}
In terms of local topology, models can be compared by their accuracy in mapping the fitness effects of single-step mutations. A primary benchmark is the replication of experimental fitness values, as illustrated in Fig.~\ref{fig:fig2}b, which shows the model prediction versus the protein fitness in the DMS of the TEM-1 $\beta$-lactamase gene~\cite{firnberg2014comprehensive}.

As described above, DMS provides an empirical measurement of the mutants' fitness, as determined by survival under antibiotic selection, of single-point mutants relative to a reference sequence.
Generative models can provide in-silico predictions of these mutational effects across various functional backgrounds~\cite{figliuzzi2016coevolutionary, hopf2017mutation} by simply computing the difference in score of the mutant sequence $\mathbf{s}^{\text{mut}}$ relative to the reference one $\mathbf{s}^{\text{wt}}$, namely

\begin{equation}
    \Delta \score^{\text{mut}} = \log \left( \frac{P(\mathbf{s}^{\text{mut}})}{P(\mathbf{s}^{\text{wt}})} \right) = \score(\mathbf{s}^{\text{mut}}) - \score(\mathbf{s}^{\text{wt}}).
    \label{eq:deltaE}
\end{equation}
It is worth noting that in the case of a single amino acid substitution in position $i$ between amino acids $a$ and $b$, equation \eqref{eq:deltaE} simplifies to:
\begin{equation}
\begin{aligned}
\Delta \score_i(a \rightarrow b)
&= h_i(b) - h_i(a) \\
&\quad + \sum_{j\neq i} \left [J_{ij}(b, s_j) -J_{ij}(a, s_j) \right]
\label{eq:delta_score}
\end{aligned}
\end{equation}
The number of terms in this sum is of order $O(2L)$ where $L$ is the length of the aligned protein sequence. As we will see later, this property makes generating stepwise evolutionary trajectories very efficient computationally.  
It has been extensively shown that this measure correlates well with experimental fitness effects~\cite{figliuzzi2016coevolutionary, riesselman2018deep, trinquier2021efficient}. 
Moreover, when the model is trained on a specific protein family, mutational effects can be compared and predicted across highly diverged homologs. 
By comparing model prediction versus the DMS of two antibiotic resistance genes from the B1 metallo-$\beta$-lactamase family (VIM-2 and NDM-1), Refs.~\cite{chen2023understanding, tsishyn2026structure} showed that DCA models capture this context dependence of mutations with reasonable accuracy. 

This behavior follows directly from the epistatic terms in Eq.~\eqref{eq:delta_score}, which make DCA predictions specific to the sequence background in which a mutation is evaluated.
A few studies have also examined the relationship between DCA-predicted mutational effects and specific protein properties, such as folding stability~\cite{levy2017potts}, melting temperature~\cite{morcos2014coevolutionary, flynn2017inference}, and catalytic activity. In the case of enzymes in particular, it was shown~\cite{xie2022enhancing} that DCA model scores correlate differently with catalysis and stability depending on the region considered, with distinct associations for the active-site region and for more distal parts of the protein. 

The ability of DCA models to predict mutational effects has also been tested for viral proteins. 
By training a model on a set of protein sequences obtained from patients, it has been shown for specific proteins of HIV~\cite{ferguson2013translating, mann2014fitness} and HCV~\cite{hart2015empirical} that \textit{in vivo} fitness is linearly related to the model score.
This result highlights the relevance of the model also in the clinical setting, where it can be helpful in understanding and predicting pathogen evolution. We will discuss this matter more in depth in \autoref{chap:res}.

As shown in Fig.~\ref{fig:fig2}b in the case of the DMS of the TEM-1 gene~\cite{firnberg2014comprehensive}, while the model shows good performance in identifying deleterious mutations, its predictive power is reduced for neutral or beneficial ones. 
This trend has been consistently observed across different systems and may stem from the way the model is trained. 
Because DCA is inferred from a finite set of natural sequences, mutations that are not observed in the training data are often assigned a deleterious effect. 
However, some of these mutations may be absent from natural sequences simply due to limited sampling, and could in fact be neutral or even beneficial.

\subsubsection*{Neutral drift experiments}
Another informative class of benchmarks for generative models is provided by sequence data obtained in neutral drift experiments, in which proteins accumulate multiple mutations with sufficient selection pressure to preserve function while maximizing genotypic diversity. 
These experiments usually start from a single wildtype sequence and undergo multiple rounds of mutation and selection. 
In this regime, sequences can move away from the reference protein while remaining confined to a functional region of sequence space. 
This makes such experiments especially relevant for testing whether generative landscape models capture well the neutral neighborhoods on which much protein evolution is expected to occur. Functional sequences with comparable function to the wildtype are expected to yield comparable statistical scores despite the accumulation of mutations.

Several studies have generated large sequencing datasets in this setting, including experiments using wildtypes from class A serine-$\beta$-lactamases~\cite{fantini2020protein, stiffler2020protein}, AAC(6) aminoglycoside N-acetyltransferases \cite{stiffler2020protein} and, more recently, B1 metallo-$\beta$-lactamases ~\cite{erdougan2023neutral}. 
These datasets are particularly valuable because they contain populations of sequences carrying multiple, likely epistatic, mutations, and therefore probe the model beyond the single-mutant regime typically accessible through DMS.

A central observation emerging from these experiments confirms that DCA scores remain close to the wild-type value even as mutations accumulate along the evolutionary trajectory, see for example \cite{bisardi2022modeling}. 
The model correctly places these variants in a relatively flat and functional region of the inferred landscape, rather than predicting a systematic loss of fitness with increasing mutational distance. 

This behavior is illustrated in Fig.~\ref{fig:fig2}c for two trajectories of the neutral drift experiment in~\cite{erdougan2023neutral}, $T10$ and $T1000$. 
Each branch corresponds to an independent 100-round long-term evolution experiment performed at either $10\, \mu g/ml$ (weak selection) or $1000\, \mu g/ml$ (strong selection) of ampicillin concentration for the clinically relevant antibiotic resistance gene VIM-2. The bright green trajectory represents a previous adaptive walk (AW) trajectory of $18$ rounds aimed at increasing the resistance of the wild-type VIM-2 gene.
Although the DCA model is trained only on natural homologous sequences very distant from the gene in question and has no direct access to the experimental data, it assigns higher scores to sequences selected at higher antibiotic concentrations. We also note that while only a modest score improvement is observed for the AW trajectory, most of the mutations that were fixed in that phase were located on the signal peptide, which is not modeled by DCA in this figure. 
\subsubsection*{Natural mutability}

Predictions from generative models on local features of fitness landscapes can also be compared with natural amino acid variability data in a specific species, as opposed to experimental ones such as DMS.
DCA models trained on pre-pandemic data (homologs across distinct coronaviruses) have been successfully employed to identify mutable sites and to predict future viral variants and viral escape potential in SARS-CoV-2~\cite{rodriguez2022epistatic}. 
More precisely, each site in the receptor binding domain of the spike protein was assigned a mutational score by the model, given as the average mutational effect $\langle\Delta \score^{\text{mut}} \rangle$ over all possible 19 amino-acid mutations.
This metric was used to predict the mutability of protein residues obtained from databases of SARS-CoV-2 genomes. 
Interestingly, these predictions were increasingly confirmed as more mutations and amino acid variability emerged during the pandemic.

Further validation of local accuracy is provided by the analysis of \textit{Escherichia coli} genomes, where DCA models trained on distant homologs accurately predicted the distribution of segregating polymorphisms~\cite{vigue2023predicting}. 
This study highlights the power of encoding co-evolutionary signals through DCA; by comparing epistatic models to profile models, it was shown that $30\%-50\%$ of sites are strongly mutationally constrained by their specific genetic context, i.e., the amino acids present in other positions. 

The importance of the sequence context in dictating the evolution of sites is captured by the model and can be quantified by comparing the site-specific global and local mutability. 
The global mutability (also called context-independent entropy) is obtained by computing the empirical frequency $f_i(a)$ of occurrence of amino acid $a$ on site $i$ across the multiple sequence alignment of diverged homologs. 
This frequency is used to compute a standard Shannon entropy as
\begin{equation}
    \text{CIE}_i = -\sum_{a=1}^{21} f_i(a) \log_2 f_i(a).
    \label{eq:cie}
\end{equation}
This metric effectively measures the global site mutability, i.e., the extent to which a site is variable or conserved across the input MSA, effectively representing the ``average'' constraint that a site experiences within a protein family. 
In contrast, the local mutability (also called context-dependent entropy) is a  metric defined for site $i$ within a specific sequence $\mathbf{s}$ as
\begin{equation}
    \text{CDE}_i(\mathbf{s}) = -\sum_{a=1}^{21} P(s_i = a | \mathbf{s}_{\setminus i}) \log_2 P(s_i = a | \mathbf{s}_{\setminus i}).   
    \label{eq:cde}
\end{equation}
Here, the quantity $P(s_i=a|\mathbf{s}_{\setminus{i}})$ is the probability assigned by the model to the presence of amino acid $a$ at site $i$, \textit{given} the rest of the sequence $\mathbf{s}_{\setminus{i}}$.
The local mutability (CDE) measures how much a site is mutable when considering a local neighborhood of a sequence.
By comparing these two metrics, one can quantify how much the specific sequence background restricts the available mutational space. 
While global mutability (CIE) provides a global view of conservation, local mutability (CDE) reveals that many sites appearing variable across a family are actually highly constrained when the specific interactions with neighboring residues in the current genetic background are taken into account. 

These quantities have been used \cite{vigue2022deciphering} to characterize the variability of amino-acid sites, accurately distinguishing conserved from polymorphic positions and quantifying the extent to which genetic context constrains sequence diversity. 
In particular, the comparison between local and global mutability provides a direct measure of evolutionary contingency, revealing how epistatic interactions reduce the set of tolerated amino acids in a given sequence background relative to what would be expected from global conservation patterns encoded in the protein family alignment.
This demonstrates how DCA can be used to investigate recent selection events and quantify how mutational preferences evolve across different genetic backgrounds within a species. 

\subsubsection*{Contact prediction}

Historically, DCA has been extensively used for residue contact prediction. 
By disentangling direct from indirect correlations of pairwise amino acid statistics in multiple sequence alignments, DCA identifies pairs of residues whose inferred couplings strongly correlate with the spatial proximity of residues in the three-dimensional structure~\cite{weigt2009identification, morcos2011direct, ekeberg2013improved,morcos2014direct}. 
This correspondence implies that the inferred $J_{ij}$ parameters are directly interpretable in structural terms, as they capture effective residue–residue interactions. 
These predictions have proven sufficiently accurate to enable de novo structure determination for proteins with large enough sequence families \cite{ovchinnikov2017protein}, and have been successfully extended to the discovery of alternative conformations \cite{morcos2013coevolutionary} and protein–protein interactions \cite{gueudre2016simultaneous}.
Overall, these results demonstrate that co-evolutionary signals captured by DCA encode global structural constraints in protein fitness landscapes solely based on statistical modeling of sequence data.

\subsubsection*{De novo sequence generation}

Beyond the prediction of local mutational effects and global structural constraints, an additional question is whether the sequence landscapes inferred by these models are truly generative. More specifically, one can ask whether sampling from them reproduces the statistics of natural sequences and assigns high probability $P(\mathbf{s})$ to functional regions of sequence space from which novel viable sequences can be drawn.
In the context of fitness landscapes, this amounts to asking whether sequences with high scores according to the model correspond to sequences with biologically meaningful properties.

Computationally, one can verify that models reproduce the statistical properties of the natural sequences they are trained on~\cite{figliuzzi2018pairwise, trinquier2021efficient, mcgee2021generative}, which serves as a primary indicator of generative capacity~\cite{socolich2005evolutionary}.
In fact, samples from DCA models reproduce by construction the amino-acid abundances $f_i(a)$ and the pairwise amino-acid co-occurrences $f_{ij}(a,b)$ of the protein family alignment used to train the model.

Furthermore, sequences generated by these models can be experimentally tested for function. 
Studies have shown that DCA models, which explicitly capture residue-residue correlations~\cite{park2024simplicity}, can successfully design active enzymes that fold into their native structures~\cite{tian2018co, russ2020evolution, tian2022design, lambert_exploring_2025, fernandez2025designing}. 
Particularly, the authors of Refs.~\cite{russ2020evolution, netti2026expanding} experimentally tested DCA-designed chorismate mutase enzymes through in vivo high-throughput complementation assays.
The study analyzed large libraries of synthetic proteins with substantial sequence diversity. 
Interestingly, sequences with low DCA score are inactive, whereas a significant fraction ($\sim 30\%$) of high score samples from the DCA-landscape exhibit catalytic activity comparable to natural enzymes, despite being highly divergent from any known sequence. 
Fig.~\ref{fig:fig2}d presents a plot of relative enrichment against DCA model score for Chorismate Mutases from Ref.~\cite{netti2026expanding}. The fraction of active (blue) sequences is directly correlated with their score.
 
Moreover, promising results have also been obtained by using DCA in the generation of functional rybozimes~\cite{Calvanese_2024, lambert_exploring_2025, Calvanese2025integrating}.
Indeed, DCA-based models, combined with structural constraints, were shown \cite{lambert_exploring_2025} to generate functional self-splicing ribozymes at mutational distances from a known wild-type far beyond what is accessible through random mutagenesis or deep mutational scanning. 
In particular, while random mutations lead to a rapid loss of activity after only $\sim 10$–$15$ substitutions, a fraction of DCA-generated sequences retains catalytic activity up to $\sim60$ mutations from a reference sequence of 197 nucleotides, demonstrating a remarkable ability to extrapolate functional constraints across sequence space.

Similar generative successes have been achieved with architectures capturing higher-order correlations, such as Restricted Boltzmann Machines and deep auto-regressive models~\cite{madani_large_2023, sevgen2025prot}. While deeper architectures seem to have better generative capabilities, the results depend strongly on the protein family under consideration, training quality, pre-processing of the alignment, etc. 
The ability of these models to propose novel, functional sequences far from the training data, while maintaining the statistical signatures and fold-stability of natural families~\cite{figliuzzi2018pairwise, russ2020evolution, lambert_exploring_2025}, strongly suggests that they capture the underlying constraints of the fitness landscape. 

This dual success in local mutational prediction and global sequence generation justifies, in our opinion, the use of these landscapes as a basis to model protein evolutionary dynamics. 
In the following section, we will show how we can use DCA models to produce realistic evolutionary trajectories in sequence space.

\section{Modeling the evolutionary dynamics}
\label{chap:evodin}

\begin{figure*}[htbp]
    \centering
    \includegraphics[width=0.95\textwidth]{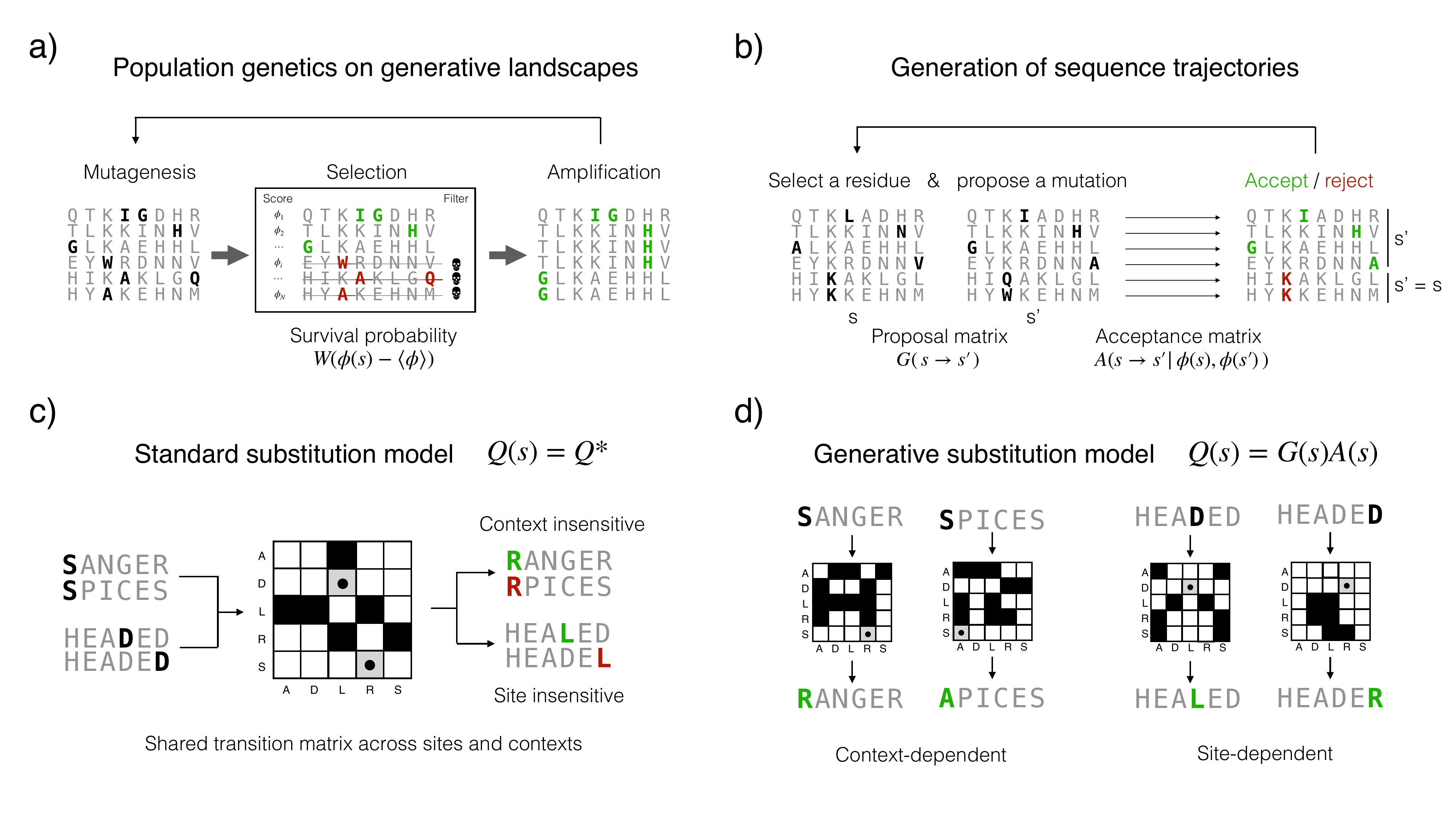}
    \caption{\textbf{Population genetics and substitution models on generative landscapes.} 
    (a) Schematic representation of a population genetics cycle on a sequence landscape: mutagenesis introduces diversity, selection filters sequences based on their generative score $\score(\mathbf{s})$, and amplification propagates functional variants. 
    The survival probability is determined by the fitness (score) of the sequence relative to the population mean, $W(\score(\mathbf{s})-\langle \score \rangle)$. 
    (b) Generation of sequence trajectories via a Markov chain Monte Carlo approach in the context of GSMs.
    At each step of the sampling procedure, a residue is selected, and a mutation is proposed according to the proposal matrix $G(\mathbf{s} \rightarrow \mathbf{s}')$; 
    the transition from the original sequence and novel sequence is accepted or rejected according to the acceptance matrix $A(\mathbf{s} \rightarrow \mathbf{s}' | \score(\mathbf{s}), \score(\mathbf{s}'))$ which is derived from the underlying sequence landscape. 
    (c-d) Comparison between a standard substitution model (site-insensitive) and a generative substitution model $Q(\mathbf{s}) = G(\mathbf{s})A(\mathbf{s})$ (context-dependent). 
    While standard models assume a shared transition matrix across sites and contexts, generative models account for the sequence context, allowing for site- and context-dependent substitution probabilities that reflect the underlying fitness landscape.}
    \label{fig:fig3}
\end{figure*}

Having shown that generative sequence landscapes can capture experimentally and evolutionarily relevant features of protein fitness across multiple divergence scales, the next step is to ask how evolution can be modeled on such landscapes. In other words, once a sequence-dependent fitness proxy has been inferred, one still needs a dynamical framework that specifies how populations or lineages move through sequence space over time.

Two major frameworks are commonly used to model protein evolution: population genetics and substitution models. While both describe evolutionary change, they differ primarily in the level of detail at which the evolutionary process is represented, particularly regarding the treatment of genetic diversity within populations.
Population genetics describes evolution by explicitly modeling the frequencies of different protein variants within a population. In this framework, multiple variants can coexist simultaneously and compete, making it particularly suitable for studying short-term, intra-species evolution in populations of microorganisms such as bacteria or yeast. 

The population can be viewed as a cloud of diverse sequences moving through the fitness landscape under the combined action of: (\textit{i}) mutation, which introduces new variants; (\textit{ii}) selection, which favors variants with higher fitness; and (\textit{iii}) genetic drift, which generates stochastic fluctuations in variant frequencies.

Such dynamics can be reproduced experimentally in highly controlled laboratory evolution experiments, where populations undergo repeated cycles of mutation and selection. By varying mutation rates and the strength of selection, these experiments can probe different evolutionary regimes~\cite{zheng2020selection}. In particular, experiments performed under strong, often progressively increasing, selection pressures are commonly referred to as directed evolution~\cite{arnold1998design}, whereas those performed under weak or absent selection are typically described as neutral or genetic drift experiments~\cite{bershtein2008intense, erdougan2023neutral}.

When the mutation rate becomes sufficiently low relative to the fixation timescale, a different approximation can be employed. In this weak-mutation regime, populations remain approximately monomorphic because new mutations arise only rarely and typically reach either extinction or fixation before subsequent mutations appear. Under these conditions, the full population dynamics can be simplified into an origin-fixation process~\cite{sella2005, mccandlish2014modeling}, in which evolution proceeds through a sequence of discrete substitution events. Rather than tracking a distribution of variants, the system is represented by a single sequence that occasionally changes when a mutation successfully fixes in the population. This approximation forms the basis of many fitness-informed substitution models \cite{halpern1998evolutionary, mccandlish2014modeling} and provides a natural framework for describing long-time evolutionary trajectories on generative sequence landscapes.

\subsection*{Population Genetics}
To model evolution on short and intermediate time-scales, population genetics provides a fine-grained framework. 
This approach is classically the most biologically realistic as it explicitly accounts for the stochastic nature of mutation and the deterministic pressure of selection within a single species. 
Historically, evolution at this scale is characterized by the change in allele frequencies over generations, traditionally modeled through the discrete dynamics of the Wright-Fisher or Moran processes~\cite{fisher1999genetical, wright1931evolution, moran1958random}.
These models are specifically designed to capture the competitive dynamics between individuals and the impact of genetic drift in finite populations.

These classical models have been integrated in the context of generative sequence landscapes. 
Seminal works have incorporated the model's score as a proxy for selective pressure inside traditional population genetics frameworks to study pathogen evolution ~\cite{shekhar2013spin, barton2016relative, hart2018computational} (see also~\cite{Choudhuri2026} for a recent review) and to model experimental evolution \cite{Dibari2026modeling, sesta2021amala}.
In this setting, evolutionary simulations are usually formulated through three idealized steps that mimic the basic ingredients of natural evolution and are also well suited to reproduce simplified experimental evolution protocols.
The steps of the dynamics are discrete, involving mutation, selection, and amplification (see Fig.~\ref{fig:fig3}a for a schematic representation):
\begin{itemize}
    \item Mutation: A population of sequences undergoes stochastic variation. 
    This is modeled by a mutational process where each site (nucleotide or amino acid) has a chance to change based on a defined mutation rate.
    \item Selection: Individual sequences are filtered based on their fitness. 
    In the generative sequence landscape framework, the survival probability of a clone of a variant sequence depends on the generative landscape model score defined for that genotype. 
    Different survival probabilities have been defined.  
    This step significantly reduces the population size.
    \item Amplification: To simulate the restoration of the population size for the next round, the survivors are sampled with replacements.
\end{itemize}
As mentioned before, the connection between the generative inferred sequence landscape and Wright-Fisher dynamics is established through the selection step. 
While specific modeling choices may vary—either by explicitly calculating the selection coefficient for each variant or by directly modeling a survival probability—the underlying objective remains the same.
In most frameworks, the survival probability $W(\mathbf{s})$ is expressed as a logistic function of the model score difference as
\begin{equation}
W(\mathbf{s}) = \frac{1}{1 + e^{\beta \left(\score[\mathbf{s}] -\overline{\score[\mathbf{s}]}\right)}},
\end{equation}
where $\score[\mathbf{s}]$ represents the model score of a given sequence, $\overline{\score[\mathbf{s}]}$ is the population average score, and $\beta$ acts as a proxy for selection strength.
Moreover, the model score function can be modified to account for the influence of the immune system in the case of evolution of viral proteins~\cite{barton2016relative, hart2018computational,Huot2025generative,Huot2025constrained}. 

While the population genetic approach effectively captures the path-dependent trajectories and historical contingency of specific lineages, its reliance on experimental parameters makes it less suited for exploring global sequence space. 
Unlike models designed for macro-evolutionary timescales, these simulations track real-time adaptation rather than sampling the full landscape as shown by a recent study \cite{Dibari2026modeling}.
This inherently implies that these evolutionary dynamics are not guaranteed to be generative: as steps accumulate, the samples from the model do not reproduce the protein family statistics, and no experimental test over sequences generated within this framework has been carried out.

\subsection*{Generative Substitution Models}
An alternative to population-genetic simulations is to model evolution as a sequence of substitutions occurring along a single lineage. This perspective is particularly useful in light of what we have previously discussed about generative sequence landscapes. If sequences sampled from a well-inferred landscape are experimentally functional, then a dynamics that explores this landscape while introducing biologically plausible mutations may generate meaningful evolutionary trajectories. We refer to this class of approaches as Generative Substitution Models (GSMs).

In contrast to population-genetic simulations, GSMs do not track a distribution of variants within a population. Instead, they define a stochastic substitution process on an inferred sequence landscape. The transition rules are constructed so that the dynamics explores the landscape while preserving the statistical features of the sequence in the training set. As discussed at the end of this section, this places GSMs between standard models and mutation-selection models. Like fitness-informed mutation-selection models, they use an explicit sequence landscape that assigns a score to each genotype. However, like more phenomenological substitution models, their transition probabilities are specified directly at the level of the Markov process, rather than being derived from an explicit calculation of fixation probabilities.

The core of this framework is a Markov Chain Monte Carlo (MCMC) dynamics. MCMC simulates the trajectory of a single lineage as a stochastic walk through sequence space by iterating over single mutation steps. The Markov property implies that the probability of the next sequence depends only on the current one.

The simulation process begins with a sequence $\mathbf{s}$ and moves to a neighboring state $\mathbf{s}'$ with a transition probability $Q(\mathbf{s} \to \mathbf{s}') = G(\mathbf{s} \to \mathbf{s}')A(\mathbf{s} \to \mathbf{s}')$. Each move can be decomposed into two different steps, corresponding to two distinct matrices:
\begin{itemize}
\item the proposal matrix $G(\mathbf{s} \to \mathbf{s}')$ models the probability of attempting a specific mutation. 
This type of matrix can be defined at different biological levels, from transitions in amino acid space to moves in nucleotide or codon space to account for the structure of the genetic code and mutational accessibility.
Broadly speaking, $G$ can be interpreted as a proxy of the mutational process.
\item the acceptance matrix $A(\mathbf{s} \to \mathbf{s}')$ that, once a move is proposed, determines whether the transition occurs or not based on the relative probability of $\textbf{s}'$ with respect to $\mathbf{s}$.
Indeed, $A$ can be viewed as a proxy of selection.
\end{itemize}
The acceptance matrix employs specific criteria that weight the move according to the model scores $\score(\mathbf{s})$ and $\score(\mathbf{s}')$ defined by the landscape. 
More specifically, a positive change in model score in Eq.~\eqref{eq:deltaE} is considered to be a beneficial mutation, whereas a negative change corresponds to a deleterious one.
For example, the Metropolis acceptance rule is defined by 

\begin{equation}
    A(\mathbf{s} \to \mathbf{s}') = \min \left( 1, e^{\beta [\score(\mathbf{s}') - \score(\mathbf{s})]} \right) 
\end{equation}
where $\beta$ is a scaling factor that determines the selection strength.
The proposal and acceptance steps are then iteratively repeated as represented in Fig.~\ref{fig:fig3}b. 
For technical details on the specific different implementations, we refer the reader to Refs.~\cite{de2020epistatic, bisardi2022modeling, biswas2024kinetic,  DiBari2024, Dibari2026modeling}.

Traditionally, MCMC sampling is used as a statistical tool to generate high-probability data points from complex probability distributions. 
In the context of generative sequence design, for example, it was employed to sample sequence candidates from inferred landscapes for subsequent experimental validation~\cite{russ2020evolution, tian2022design, netti2026expanding}. 

However, once these inferred landscapes were shown to reproduce well the landscape around protein sequence at short distances ($\sim10\%$ of divergence) as well as capturing global features of those landscapes (by generating functional sequences at more than $\sim 70 \%$ of divergence), as discussed in \autoref{chap:genseq}, it became natural to ask whether MCMC could be interpreted more directly as a model of evolutionary dynamics, rather than only as a way to obtain equilibrium samples. 

Under this interpretation, the intermediate states visited by the Markov chain can be viewed as plausible evolutionary intermediates rather than merely as computational steps toward equilibrium. 
Supporting this view, a landmark study~\cite{alvarez2024vivo} - that we will discuss in more detail in \autoref{chap:res} - experimentally tested $\beta$-lactamase sequences sampled at different stages of a MCMC evolutionary simulation and found them to be active against ampicillin \textit{in vivo}.

It is important to note that the MCMC dynamics underlying GSM is based on a mathematical guarantee, known as detailed balance
\begin{equation}
    P(\mathbf{s})Q(\mathbf{s} \to \mathbf{s}') = P(\mathbf{s}')Q(\mathbf{s}' \to \mathbf{s})
    \label{eq:detailed_balance}
\end{equation}
that is to say, the probability of being at state $\mathbf{s}$ and transitioning to $\mathbf{s'}$ is the same as being at $\mathbf{s'}$ and transitioning to $\mathbf{s}$, i.e. the dynamics is time reversible.
This property ensures that long enough substitution trajectories eventually converge to a stationary regime in which they faithfully sample the regions of sequence space spanned by the training set. Although any valid MCMC scheme satisfying detailed balance converges to the same equilibrium distribution, the specific choice of the proposal and acceptance matrices remains crucial for the realism of the dynamics at short timescales. 
In particular, the proposal matrix determines which mutations are locally accessible during evolution. 
For example, codon-level moves preserve the connectivity imposed by the genetic code in amino-acid space and therefore more faithfully represent which variants can be reached in the early stages of evolution~\cite{GONZALEZ2019fitness, gunnarsson2023predicting, Dibari2026modeling}.

A limitation of the vanilla GSM formulation is that it ignores phylogenetic constraints, since evolutionary trajectories are independent from each other and no historical correlations are imposed. 
To overcome this drawback, a tree-based variant (treeMCMC) can be implemented~\cite{Dibari2026modeling}, in which evolution is simulated along an inferred phylogenetic tree. 
Sequences evolve from the root to the leaves via MCMC dynamics, with the number of steps along each branch proportional to its length. 
This approach imposes historical constraints, introduces correlations between sequences, partially accounts for the time-ordering of substitutions, and incorporates experimental phylogenetic information into the simulation framework, thereby granting greater biological plausibility. 

Before moving to the review of results achieved by generative protein evolution models in \autoref{chap:res}, we want to briefly discuss how other classes of substitution models compare to the MCMC approach to evolution described in this section. In classical phylogenetic substitution models~\cite{pupko:hal-02535102}, the central object is a transition or rate matrix that specifies the probability, or instantaneous rate, of replacing one character state with another. This matrix is usually inferred by maximizing the likelihood of an evolutionary model on sequence data and can then be used to describe sequence evolution along a phylogeny or to simulate evolutionary trajectories. In their simplest and most widely used forms, these models are site-independent: the same substitution process is applied across sites, possibly with site-specific rate variation, but without making the substitution probability depend explicitly on the full sequence context.

Fitness-informed mutation-selection models provide a more mechanistic alternative. In these models, the substitution process is connected to an explicit fitness landscape, and transition rates can be expressed in terms of an underlying mutation process and the probability that a new variant fixes in the population~\cite{halpern1998evolutionary,mccandlish2014modeling}. Related ideas have also been developed in the context of biophysical models of protein evolution~\cite{liberles2012interface,shah2015contingency,echave2017biophysical,bastolla2017evolution}.

GSMs share features with both classes of models. Like mutation-selection models, they rely on an explicit sequence landscape that assigns a score to each genotype and can therefore be interpreted as a proxy for fitness. However, in the GSM formalism considered here, transition probabilities are generally not derived from an explicit population-genetic calculation of mutation and fixation probabilities. Instead, as in more phenomenological substitution models, the transition structure is specified directly at the level of the Markov process. The distinctive feature of GSMs is that this process is built on a generative model already inferred to reproduce the static distribution of the training family, and the transition rules are chosen to preserve this distribution. While the principles of this approach are not new \cite{halpern1998evolutionary, sella2005}, previous models rarely considered interaction between sites explicitly, and never on generative sequence landscapes.

This difference becomes clear when comparing local transition matrices, as schematized in Fig.~\ref{fig:fig3}c-d. In a standard site-independent substitution model, the same transition matrix is applied across sites and sequence contexts. In a GSM, by contrast, one can define an effective local transition matrix after fixing a sequence context. Given a current sequence $\mathbf{s}$, the probability of mutating site $i$ from residue $a$ to residue $b$ is determined by the proposal and acceptance terms, and therefore by the score difference between $\mathbf{s}$ and the corresponding mutant sequence. Since this score difference is computed from an epistatic generative landscape, the resulting local transition matrix depends both on the site being mutated and on the rest of the sequence background.

We note that most of the models considered in this review are formulated with a discrete-time formalism. However, it is possible to construct a GSM defined on a continuous time~\cite{pagnani2025generative}.
Its dynamics converge to the prescribed equilibrium distribution 
($P(\mathbf{s})$) and the model is defined by a sequence-to-sequence transition rate matrix \(Q(\textbf{s)}\) constrained to satisfy detailed balance.
Mutation events and times are selected through a Gillespie algorithm~\cite{Gillespie1977Exact}. 
For computational tractability, only single–amino-acid (or nucleotide) substitutions are allowed, so that each sequence has $L(q-1)$ possible moves. 
Hence, this framework never requires explicit construction of the full \(q^L \times q^L\) matrix that would formally describe the MCMC dynamics in sequence space.

\begin{table*}[t]
\centering
\caption{Summary of the key studies discussed in this review that use generative sequence landscapes as proxies for modeling protein evolution.
For each study, the table reports the evolutionary model, the biological system considered and a brief description of the results. Abbreviations: GSM, Generative Substitution Model; PopGen, generative population genetics.
}
\footnotesize
\label{tab:evolution_gfl_summary}
\setlength{\tabcolsep}{6pt}
\renewcommand{\arraystretch}{1.8}

\begin{tabular}{p{1.3cm} p{3.5cm} p{8cm} p{3.2cm}}

\toprule
{\small \textbf{Model}} & {\small \textbf{Gene family}} & {\small \textbf{Results}} & {\small \textbf{Refs}} \\
\midrule

PopGen &
HIV-1 p17 matrix protein  &
Showed why DCA models trained on patient-derived viral sequences reflect the correct rank order of mutant viral fitness.&
\citet{shekhar2013spin}, \newline (2013) \\

PopGen &
HIV-1 all genes &
Quantified the predictability of intra-host HIV evolution and immune escape pathways. &
\citet{barton2016relative}, \newline (2016) \\

PopGen &
HCV NS5B polymerase &
Computationally designed HCV vaccines leveraging simulations on inferred viral fitness landscapes. &
\citet{hart2018computational}, \newline (2018) \\

PopGen &
HCV E2 glycoprotein  &
Developed a predictive in-silico evolutionary model for targeted vaccine design. &
\citet{quadeer2019identifying}, \newline (2019) \\

GSM &
Multiple protein families &
Unified neutral protein-evolution models and linked evolutionary Stokes-shifted residues to biologically relevant regions. &
\citet{de2020epistatic}, \newline (2020) \\

GSM &
ClassA $\beta$-lactamase \& \newline AAC6 acetyltransferase &
Characterized sequence space exploration and its impact on the emergence of epistatic signals during neutral evolution experiments.  &
\citet{bisardi2022modeling}, \newline (2022) \\

GSM &
HIV protease and reverse trascriptase &
Elucidated the role of contingency and entrenchment in the fixation of resistance mutations. &
\citet{choudhuri2022contingency}, \newline (2022) \\

PopGen &
HCV E2 glycoprotein  &
Conducted a comparative analysis of the fitness landscapes between HCV subtypes 1a and 1b linking it to disease chronicity. &
\citet{zhang2022evolutionary}, \newline (2022) \\

PopGen &
HCV E1/E2 complex &
Showed that fitness-compensating E1 mutations accelerate viral escape from E2-targeting antibodies. &
\citet{zhang2023hcv}, \newline (2023) \\

PopGen &
HCV NS3 protease &
Demonstrated how compensatory epistatic interactions facilitate rapid escape from drug selection pressure. &
\citet{zhang2023direct}, \newline (2023) \\

GSM &
Multiple protein families &
Characterized the long-term timescales of epistasis and site-specific mutability during protein evolution. &
\citet{DiBari2024},\newline  (2024) \\

GSM &
HIV-1 protease, reverse transcriptase and integrase&
Developed kinetic coevolutionary models to predict multi-drug resistance mutations. &
\citet{biswas2024kinetic}, \newline (2024) \\

GSM &
ClassA $\beta$-lactamase &
Validated computational epistatic models by directly linking them to observed in vivo functional phenotypes. &
\citet{alvarez2024vivo}, \newline (2024) \\

GSM &
Multiple protein families &
Introduced continuous-time generative models to capture dynamical epistatic signatures. &
\citet{pagnani2025generative}, (2025) \\

GSM &
Multiple protein families &
Disentangled mutational fluctuations inherited from ancestral lineages from those driven by stochastic drift. &
\citet{rossi2024fluctuations}, \newline (2025) \\

GSM &
Multiple protein families &
Analyzed the strict constraints epistasis places on contingency and entrenchment in neutral evolution. &
\citet{Schmelkin2025entrenchment}, \newline (2025) \\

PopGen \& GSM &
ClassA $\beta$-lactamase \& \newline AAC6 acetyltransferase \&\newline   DHFR&
Benchmarked GSM and generative population genetics simulations directly against in-vitro evolution experiments. &
\citet{Dibari2026modeling}, \newline (2026) \\

GSM &
Multiple protein families &
Demonstrated that out-of-equilibrium protein dynamics and fluctuating selection significantly DCA inference quality. &
\citet{dietler2025out}, \newline (2026) \\

\bottomrule
\end{tabular}
\end{table*}

\section{Results \& applications}
\label{chap:res}

\begin{figure*}
    \centering
    \includegraphics[width=0.95\textwidth]{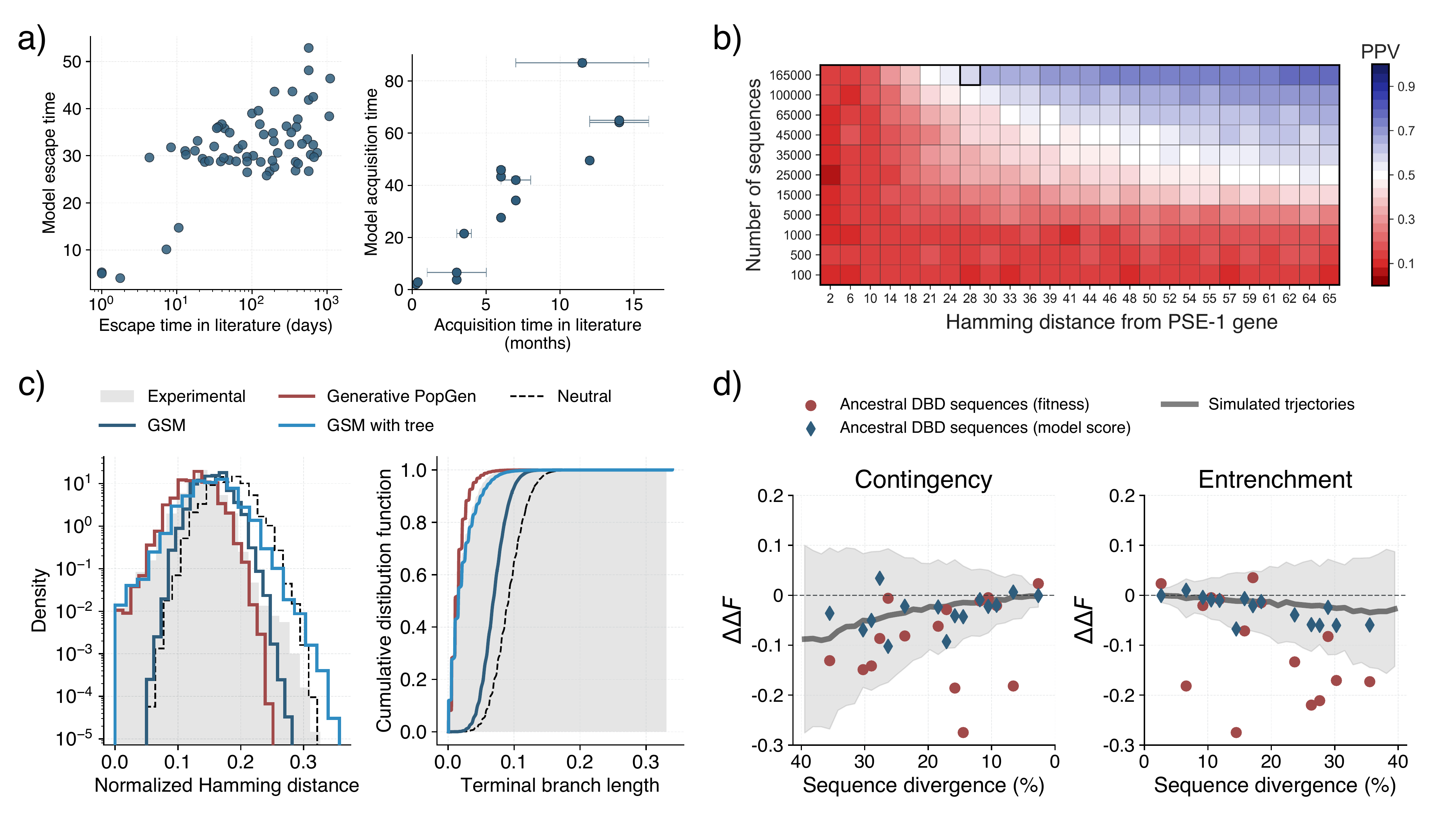}
    \caption{\textbf{Results on experimentally accessible timescales.} 
    (a)~Predicted versus reported timing of HIV-1 immune escape (left, measured in days)~\cite{barton2016relative} and acquisition of drug-resistance mutations (right, measured in months)~\cite{biswas2024kinetic}, demonstrating the accuracy of generative evolutionary simulations in modeling virus evolution.
    (b)~Positive predictive power (PPV) for structural contacts of the PSE-1 gene by DCA as a function of sequence divergence and sample size of the training MSA. Each MSA was generated by a GSM simulation of a neutral drift experiment. Blue represents high accuracy, and red represents low accuracy~\cite{bisardi2022modeling}. 
    (c)~Comparison of phylogenetic statistics of different evolutionary frameworks against experiments. Pairwise sequence distances (left) and terminal branch lengths (right) for experimental sequences (gray), GSM simulations (dark blue), GSM simulations on a tree (light blue), and generative population genetics simulations (red), illustrating the improved historical fidelity of population genetic frameworks~\cite{Dibari2026modeling}. 
    (d)~Quantification of evolutionary constraints in the DNA-Binding Domain (DBD) family: contingency (left) and entrenchment (right)~\cite{DiBari2024}. 
    Red points denote experimental DMS fitness measurements, blue diamonds represent DCA-based scores, and grey lines indicate GSM simulations with standard deviations (shaded areas). 
    $\Delta \Delta F$ indicates the shift in mutational impact over sequence divergence. 
    Figures are reproduced with permission and data from the original authors.
     }
    \label{fig:short_time}
\end{figure*}

Building on the methodologies discussed above, we now turn to the results obtained by integrating generative sequence landscapes with protein evolution models. To provide a coherent perspective across different biological and temporal regimes, we organize the discussion according to both the evolutionary scale and the nature of the phenomena under investigation. 

We begin with short-term, directly observable evolutionary dynamics, where model predictions can be quantitatively compared with clinical and experimental data. 
We then consider controlled in vitro evolution experiments, which enable a systematic benchmarking of simulations.
Next, we show that these models quantitatively reproduce key epistatic evolutionary phenomena such as contingency and entrenchment.

At the same time, simulations provide access to a much richer set of observables than those currently accessible experimentally, allowing us to probe their full dynamical structure beyond the partial snapshots provided by available data.
This also enables the use of these models in regimes that are currently difficult or impossible to access experimentally. 
In particular, they can be leveraged to study long-term effects of epistasis and statistical properties of protein evolution.

A selection of results obtained by different kinds of generative protein evolution  models is reported in Table~\ref{tab:evolution_gfl_summary}.

\subsection*{Prediction of short-term viral evolution}

Early applications combining population genetics models with DCA-inferred fitness landscapes focused on virus evolution. 
The first papers focused on identifying specific ``vulnerable sites'', suitable as antibody epitopes.
These are defined here as residues where the fitness cost of escape mutations is prohibitively high across diverse sequence backgrounds, making them optimal targets for therapeutic intervention in the HIV-1 proteome~\cite{shekhar2013spin, barton2016relative}. 
After validating the ability of the DCA model to reproduce mutational effects across several HIV protein domains, the inferred landscape—derived in this case from patient HIV sequence data—was integrated into a population genetics framework incorporating selection pressure \textit{via} immune response~\cite{barton2016relative}. 
Remarkably, this approach was able to predict with reasonable accuracy the timing of escape mutations observed in longitudinal patient data, as shown in Fig.~\ref{fig:short_time}a (left panel).

Recent advances in modeling HIV evolution \cite{biswas2024kinetic, choudhuri2026temporal} leveraged GSMs 
simulations to predict the temporal emergence of drug-resistant mutations in patients undergoing antiretroviral therapy as shown in Fig.~\ref{fig:short_time}a (right panel).
A DCA model was learned on sequences from drug-experienced patients, while the evolutionary trajectories were simulated starting from drug-naive patient sequences. 
The acquisition times of drug-resistant mutations matched literature times with very high accuracy.
It is worth mentioning that a static mutational metric, depending on the DCA model and the family-wide site conservation patterns was defined in order to predict dynamical features of the emergence of drug-resistant mutations.
 
Similar works have shown in the context of Hepatitis C Virus (HCV) that empirical landscapes could be used to guide the design of immunogens by identifying the specific fitness costs associated with escape mutations, effectively demonstrating that population genetics on DCA-derived landscapes can predict the vulnerabilities of a virus before they are exploited~\cite{hart2018computational}.
Comparable successes have been recorded in the study of the E2 glycoprotein of HCV ~\cite{quadeer2019identifying, zhang2022evolutionary} that revealed that fitness-compensating mutations in the E1 protein can accelerate escape from antibodies~\cite{zhang2023hcv, zhang2023direct}. 
More recently, RBM models have anticipated the escape potential of SARS-CoV-2 by modeling evolutionary funnels shaped by immune pressure~\cite{Huot2025generative, Huot2025constrained}. 

\subsection*{Quantitative modeling of genetic drift experiments}

One of the most interesting applications of GSM and generative population genetic models regards the modeling and simulation of neutral drift experiments. High-throughput sequence data from genetic drift experiments is unfortunately not very common, but a few interesting works sequenced thousands to hundreds of thousands of protein variants across different protein families~\cite{stiffler2020protein, fantini2020protein, dacosta2023inferring}. Those studies used respectively class A serine-$\beta$-lactamases, acetyltransferases and dihydrofolate reductases.

These experiments are performed starting from a wild-type reference of interest, with rounds of random mutagenesis and mild selection under antibiotics, resulting in a diverse set of functional sequences. 
In this context, short-time GSM simulations faithfully replicated the sequence statistics and fitness trajectories observed in the lab~\cite{sesta2021amala, bisardi2022modeling,  DiBari2024, Dibari2026modeling}. 

In Ref.~\cite{bisardi2022modeling}, the authors performed realistic simulations of genetic drift experiments using a nucleotide-based GSM.
They concentrated on data from experiments~\cite{stiffler2020protein, fantini2020protein} on the class A serine-$\beta$-lactamase family produced by rounds of random mutagenesis and antibiotic-based selection. 
Experimental sequences at the last round were used to infer structural contacts \cite{ekeberg2013improved}, giving different results in the two experimental settings. 
In Ref.~\cite{bisardi2022modeling} thousands of different experimental conditions were simulated using GSMs, finding that the divergence from the wild-type and the number of experimentally tested sequences were the key factors influencing the inference of the structural contacts, demonstrating how the different experimental conditions led to very different contact prediction accuracy as shown by Fig.~\ref{fig:short_time}b in the case of experiments evolving the PSE-1 gene from class A $\beta$-lactamase.
This study underlines how realistic simulations can provide insights that would be otherwise inaccessible under current experimental capabilities.
 
In a subsequent work~\cite{DiBari2024}, it was shown that \textit{in silico} GSM simulations correctly reproduce the residue mutability patterns found experimentally.
Indeed, the GSM was tuned to match the same experimental selective pressure and divergence time. 
Simulations correctly reproduced the effective number of tolerated amino acids for each protein site observed in the experimental sequences.

Finally, a comprehensive benchmark of generative substitution models and population dynamics against a variety of different evolution experiments~\cite{Dibari2026modeling} was recently carried out.
The study found that while GSM-based simulations reproduce general divergence patterns and the broad statistical features of the populations, it is necessary to include phylogenetic information or population genetic-like evolutionary schemes to correctly model selective sweeps and improve historical fidelity. 
The authors inferred phylogenetic trees over experimental sequences and simulated data using GSM and population genetics.
As depicted in Fig.~\ref{fig:short_time}c, GSM simulations cannot reproduce the left tail of pairwise sequence distances (left panel) nor the small terminal branch lengths of the inferred trees, whereas population genetics aligns with experimental results.
This confirms the fact that when limited to short-time scale evolution within a single species, generative population genetics frameworks are most suited as a simulation tool to mimic protein evolution.

\subsection*{Quantitative modeling of contingency and entrenchment}
A stringent validation of these frameworks lies in their ability to reproduce historically dependent constraints on evolution.
These constraints manifest when a specific mutation is either enabled by some prior mutations, termed contingency, or rendered hard to revert due to the stabilizing effect of subsequent co-evolving mutations, termed entrenchment \cite{shah2015contingency, starr2018pervasive, park2022epistatic}. 
DCA-inferred landscapes accurately predict that the historical state of a sequence determines its subsequent tolerance to mutations.

These phenomena, which can be, for example, shown by comparing DMS mutational effects from diverging homologs across reconstructed lineages, as in the steroid receptor DNA-binding domain~\cite{park2022epistatic}, were qualitatively reproduced in silico using computational approaches~\cite{DiBari2024}.

Experimental measurements of fitness effects enabled the quantification of contingency (Fig.~\ref{fig:short_time}d, left panel), which arises when the effect of a mutation depends on the genetic background in which it occurs. 
To measure this phenomenon, amino acids that first appeared at a given point along an evolutionary trajectory were introduced into ancestral sequences that predated their emergence. 
If the mutation is less favorable or even deleterious in these earlier backgrounds, this indicates that its fixation depended on preceding substitutions, revealing historical contingency. 
As sequence divergence increases, the fitness effect of the same mutation changes, providing a quantitative measure of the extent to which evolutionary history constrains future evolutionary paths.

Conversely, entrenchment (Fig.~\ref{fig:short_time}d, right panel) describes the tendency of a substitution to become increasingly difficult to reverse as subsequent mutations accumulate. 
To quantify entrenchment, amino acids that disappeared for the last time at a given point along the trajectory were reintroduced into descendant sequences. 
If the ancestral amino acid becomes progressively less tolerated in these later backgrounds, this indicates that subsequent substitutions have adapted to the derived state, thereby reinforcing its fixation. 
The observed decrease in the fitness of reversions with increasing evolutionary distance demonstrates how co-evolving mutations stabilize previously acquired substitutions and make evolutionary reversals increasingly unlikely. 

A substantial body of work has also demonstrated that GSMs can quantitatively reproduce contingency and entrenchment patterns also in HIV-1 protease and subtype B~\cite{flynn2017inference, biswas2019epistasis, choudhuri2022contingency}.

However, recently published work~\cite{Schmelkin2025entrenchment} has argued that contingency and entrenchment effects fail to manifest when simulations are conducted under neutral conditions, specifically when sequences are confined to a limited range of the model score within the landscape. Consequently, further research is required to fully elucidate the conditions under which these evolutionary phenomena emerge.


\begin{figure*}
    \centering
    \includegraphics[width=0.95\textwidth]{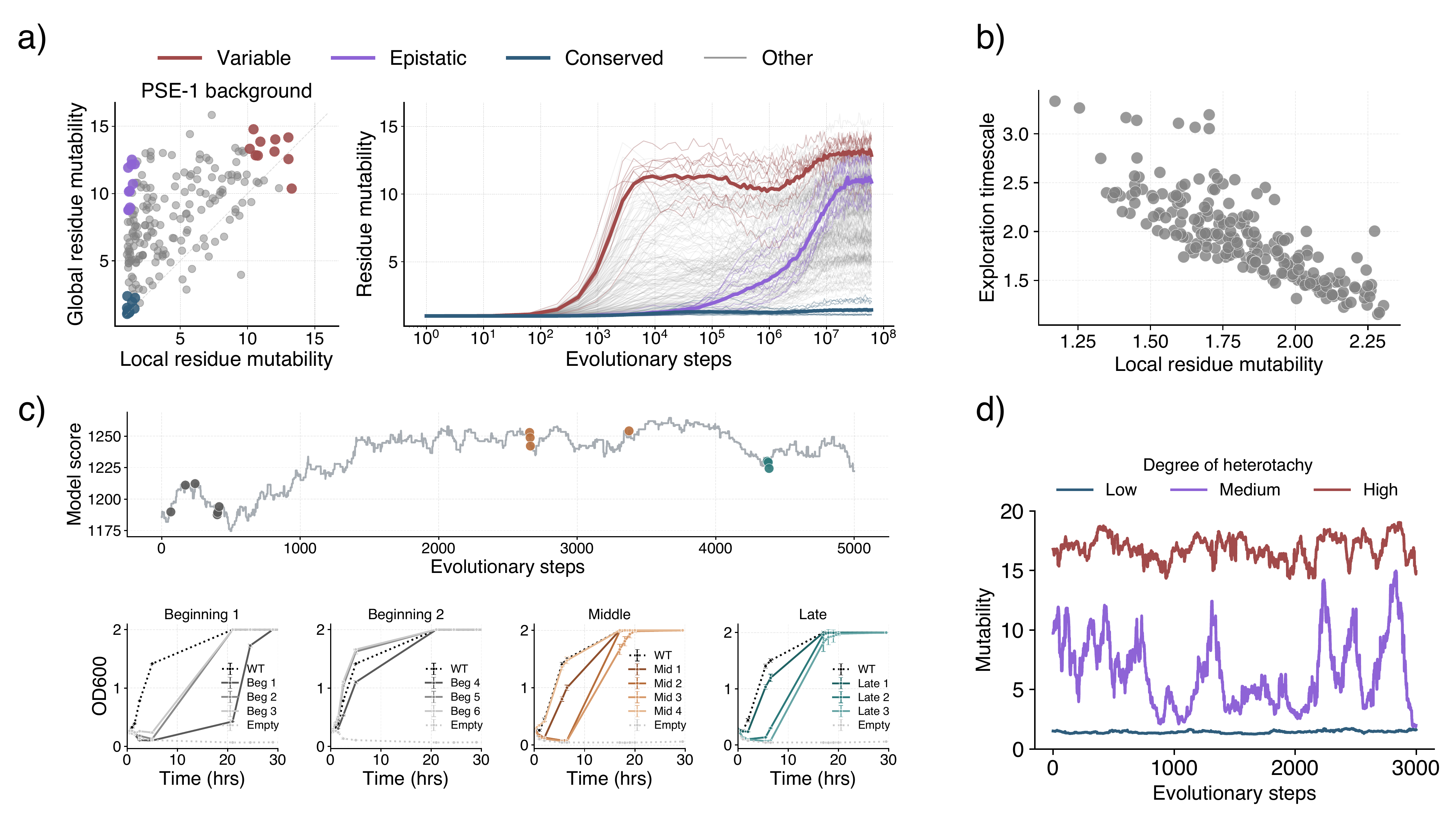}
    \caption{\textbf{Results beyond currently accessible experimental timescales.} 
    (a) In the left panel, local site mutability is compared to global site mutability for the PSE-1 gene, with residues colored by category: locally and globally mutable (red), locally and globally conserved (blue), and locally conserved but globally mutable (violet). In gray, all other residues. The right panel shows the temporal evolution of site mutability across residues during an evolutionary simulation of multiple parallel trajectories ~\cite{DiBari2024}. 
    (b) The panel displays the relationship between local sequence mutability and the time required to explore the sequence space for independent lineage GSM simulations~\cite{rossi2024fluctuations}. 
    (c) The top plot displays model scores over evolutionary steps of a single GSM trajectory starting from the TEM-1 gene, while the bottom panel presents \textit{in vivo} OD600 growth assays for wild-type (WT) and synthetic intermediates from the beginning (gray), middle (ochre), and late (aquamarine) stages of the trajectory, demonstrating that simulated trajectories maintain biological function in vivo~\cite{alvarez2024vivo}.
    (d) The effective number of amino acids shown as a function of evolutionary steps, with curves colored by the degree of heterotachy (low, medium, high)~\cite{de2020epistatic}.
     Figures are reproduced with permission and data from the original authors.
    }
    \label{fig:long_time}
\end{figure*}

\subsection*{Epistasis and time-dependence of sequence constraints}

One of the most significant advantages of GSMs is the ability to test evolutionary hypotheses that lie beyond current experimental reach. 
Since GSMs perform well on short, verifiable scales, and they are guaranteed to converge to the statistics of protein families at long time scales, they can be used to explore macro-evolutionary scenarios or out-of-equilibrium regimes currently impossible to recreate in a laboratory. 

These constraints can be used to quantify how fast evolution proceeds and to predict how likely different sites are to mutate over long timescales, by measuring how much variability is allowed in different sequence contexts~\cite{biswas2024kinetic, DiBari2024, rossi2024fluctuations}. 
In simple terms, they provide a way to estimate how “free” or “constrained” each position in a protein is as evolution unfolds.

Recent work~\cite{DiBari2024} has focused on understanding the characteristic timescales over which these constraints change, to explain why some residues remain highly conserved while others diversify more rapidly.
By comparing the local and global mutability measures in Eqs.~(\ref{eq:cie},\ref{eq:cde}), it becomes possible to disentangle two effects: the ``global'' intrinsic mutational tolerance of a site across the protein family, and the additional ``local'' constraints imposed by its specific sequence background. 
This reveals that the mutational behavior of a site is not fixed, but evolves over time together with the rest of the protein, as shown in Fig.~\ref{fig:long_time}a. 
When simulating evolution from a particular sequence, we observe two different behaviors. 
When local and global mutability of the site are high (left panel red dots) the evolutionary dynamics is quite fast (right panel, red lines), whereas when these two metrics substantially differ (left panel, violet dots) the residue needs a long time to become mutable (right panel, violet lines).
In practice, a residue that appears non-mutable in one context may later become highly mutable as surrounding positions change. 

Interestingly, the same phenomenon has been investigated and modeled with some success with a very different approach, namely biophysical protein evolution models~\cite{echave2017biophysical}.
This highlights that mutability is not an intrinsic, static property of a site, but a transient feature that depends on the protein’s current position in the fitness landscape. 
In this sense, the “mutational landscape” experienced by each site is continuously reshaped by the rest of the sequence.

This same concept has later been extended from single residues to the entire sequence. 
In~\cite{rossi2024fluctuations} it was shown that the average local mutability of each sequence of a protein family was highly correlated with the timescale that a GSM simulation needed to explore the sequence space (Fig.~\ref{fig:long_time}b). 
Broadly, the mutability induced by a certain sequence context will influence the speed at which multiple lineages evolving from that sequence will travel in the sequence space.

\subsection*{Properties of long-term protein evolution}

A fundamental consequence of context-dependent and time-varying mutability in GSMs is that substitution rates are inherently non-uniform across sites. Even if mutations occur uniformly at the microscopic level, the probability that they are accepted depends on the evolving sequence background.

GSMs naturally give rise to a broad, gamma-like distribution of substitution rates across sites, in agreement with classical observations in molecular evolution~\cite{de2020epistatic}. The development of continuous-time substitution formalisms has further refined our understanding of how these epistatic interactions among residues influence substitution rates over time~\cite{pagnani2025generative}. Indeed, GSMs predict heterotachy, where residues exhibit heterogeneous transition rates across time as a natural consequence of epistasis~\cite{de2020epistatic}, as shown in Fig.~\ref{fig:long_time}d. The presence of heterotachy indicates that a site's degree of constraint changes over time, primarily because the coevolutionary forces—or connectivities—among sequence positions shift as the sequence itself evolves.

Finally, GSM simulations have demonstrated the capacity to generate entirely synthetic evolutionary trajectories that maintain biological function, a significant advancement given that traditional sequence evolution models have historically struggled to maintain protein stability~\cite{iglesias2025empirical} and struggle to produce realistic sequences~\cite{trost2024simulations}. 
A pivotal study managed to produce functional sequences along a few GSM simulated lineages evolved from the TEM-1 $\beta$-lactamase gene, which were validated in vivo through expression in E. coli \cite{alvarez2024vivo}.
As shown in Fig.~\ref{fig:long_time}c, these variants accumulated dozens of substitutions distributed throughout the protein structure while preserving key catalytic and interaction sites. Remarkably, several evolved sequences retained wild-type activity and, in some cases, exhibited even greater antibiotic resistance than the ancestral TEM-1 enzyme. 
As shown in Fig.~\ref{fig:long_time}c, variants collected from the beginning, middle, and late stages of an evolutionary trajectory remained functional when expressed in E. coli, maintaining substantial antibiotic resistance despite extensive sequence divergence from the ancestral protein. These results demonstrate that long computational evolutionary trajectories can explore distant regions of sequence space while preserving biological function.

This capability to bridge the gap between theoretical modeling and functional protein design is further supported by recent integrative approaches. 
For instance, pseudo-likelihood-based models have demonstrated a similar capacity to navigate complex fitness landscapes while generating viable protein sequences \cite{fram2024simultaneous}. 
By proving that models derived from natural sequence alignments can yield functional variants with substantial mutational loads, these results demonstrate that the coevolutionary information captured by DCA-based models is sufficient to guide protein evolution toward stable, active regions of sequence space.

Overall, these results demonstrate that population genetics and generative substitution models can be used to successfully simulate protein evolution across a wide range of timescales and biological contexts. We stress here that since the inferred fitness landscapes are protein-family and genotype-specific, the results are biologically relevant and prone to application to concrete problems.
By quantitatively reproducing short-term dynamics, contingency, and entrenchment, while providing full access to complete evolutionary trajectories and the specific mutations that drive these effects, these models go well beyond the partial snapshots currently accessible in experiments.

Crucially, many of the features observed—such as context-dependent mutational tolerance, heterotachy, and evolutionary speed—arise naturally from the interactions encoded in the models. 
In other words, these are emergent properties: they are not imposed by hand, but instead result directly from the intrinsic structure of the evolutionary landscapes inferred by DCA. 
This observation suggests that these models not only serve as predictive tools, but might also be relevant mechanistic probes into the rules shaping protein evolution, from the micro-scale of individual residues to the macro-scale of long-term evolutionary trajectories.

\section{Perspectives}
\label{chap:persp}

Future work on the use of generative models to study protein evolution problems can proceed in many directions. 
In this last section, we discuss first some strategies for improving the accuracy of generative substitution models.
The directions of these improvements can be taken both on the generative fitness landscape side and on the algorithms used to reproduce evolution on them.
Finally, we highlight a range of applications for the generative protein evolution models described in this review.

\subsection*{Model improvement}

While generative sequence models (GSMs) have been studied for several decades, their ability to capture essential features of real evolutionary processes has long remained difficult to assess, primarily due to the lack of comprehensive and reliable experimental data. 
Recent technological advances, however, have led to a new wave of high-throughput experiments that make such validation increasingly feasible, offering the opportunity to more precisely evaluate our current understanding of protein evolution.
This process has already begun, but systematic comparisons between model predictions and experimental measurements remain crucial. 
In particular, it is important to test these models against diverse features of fitness landscapes, spanning both local properties—such as mutational effects near a reference sequence—and global properties related to the overall structure of sequence space. 
In this context, the growing availability of combinatorial mutagenesis data is especially valuable. 
Such datasets enable the exploration of fitness landscapes beyond single mutations, allowing for the validation of model predictions across multiple mutational steps and, in some cases, for a near-complete characterization of restricted regions of sequence space.

The precision of current models can be enhanced by addressing several key limitations. One priority is the incorporation of phylogenetic constraints to improve the prediction of mutational effects \cite{Khatri2025Phylogenetic}. 
Because phylogenetic correlations and other confounding evolutionary effects can generate statistical dependencies that are unrelated to the biological constraints targeted by inference methods, several studies have employed generative evolutionary simulations to assess the robustness of inference in the presence of factors such as shared ancestry, structural constraints, functional sectors, and out-of-equilibrium selection dynamics \cite{gerardos2022correlations,dietler2023impact,dietler2025out}.

Furthermore, the data these kinds of models feed on are unlabeled protein sequences that are considered functional as they are observed in nature. 
However, it is now possible to directly measure functionality (and/or fitness) of specific protein variants, unraveling the local and global structure of the fitness landscape.
Integrating such experimental measurements directly into the models greatly improves their performance and remains a critical direction to explore~\cite{barrat2016improving,Calvanese2025integrating}.

On the evolutionary simulation side, adding biologically realistic details such as better handling of indels and the introduction of mutational or codon biases would further bridge the gap between theory and nature.
It has indeed been observed that the accessibility of an amino acid from the reference nucleotide sequence that specifies a protein can affect its appearance in short-time evolutionary trajectories~\cite{gunnarsson2023predicting}.
Although this phenomenon cannot be observed if one focuses only on the amino-acid alphabet, simulations with the genetic code have been performed~\cite{DiBari2024}, and something similar was observed both in the GSM and in the PopGen models~\cite{Dibari2026modeling}.

The integration of architectures such as RBMs, VAEs, and LSTMs has fundamentally expanded our capacity to map functional diversity by capturing high-order epistasis within continuous latent spaces \cite{biswas2021low, Ziegler2023Latent, Shukla2025Thermal}. 
These models provide powerful evolutionary coordinates for de novo design and predicting viral escape \cite{Huot2025generative, sevgen2025prot}. 
A recent study has used latent representations to estimate pseudo-velocities and times of viral and ancient evolutionary trajectories \cite{Hie2022Evolutionary}.
However, naive interpolation in latent space often fails to reconstruct valid ancestral paths, revealing a mismatch between the geometry of these learned spaces and true phylogenetics~\cite{Gorstein2025Ancestral}.

\subsection*{Novel applications}


A significant hurdle for the widespread implementation of complex, site-dependent GSMs is the computational impossibility of calculating the likelihood of a sampled evolutionary trajectory. 
Traditional site-independent models are favored in phylogenetics because their likelihood computation is fast as their transition matrix is time-independent and factorizes across sites. 
In contrast, the transition matrix for GSMs is of astronomical size ($q^L \times q^L$) and is inherently sequence-dependent, meaning it cannot be precomputed. 
While progress is being made toward making these models more tractable \cite{jensen2000probabilistic, prillo2023cherryml, mathews2025importance, li2025gibbs}, practical inference remains challenging. 
Interestingly, some bioinformatics tasks can now be tackled with complex models by avoiding likelihood computations entirely \cite{arenas2022proteinevolverabc, blassel2025likelihood}.

In this same field, the development of GSMs and deep models opens several transformative applications. 
GSMs can serve as realistic forward simulators to provide ground truth sequences for benchmarking ancestral sequence reconstruction (ASR) algorithms. 
Given that neural networks can now distinguish between natural sequences and those from traditional models \cite{trost2024simulations}, GSMs offer a superior framework for validating ASR \cite{Zeinaty2026towards}. 
In this field, a transformer-based evolution simulator has appeared \cite{Koehl2026deep}, showing promising results for several phylogenetic tasks. 
In addition, thanks to GSMs, it is possible to bias ancestral sequence reconstruction to account for co-evolutionary correlations, potentially improving the accuracy and stability metrics of candidate ancestors \cite{deleonardis2024reconstruction, Zeinaty2026towards} or also reconstruct evolutionary intermediates among natural sequences~\cite{Netti2026reconstructability}.

On a close topic, substantial work has been done on studying transition path sampling, in which researchers focus on paths connecting two sequences with different characteristics \cite{mauri2023mutational, Rehan2025design, Schulte2025functional}.


Although the generation of protein sequences under selection pressure or neutral drift is, in principle, experimentally simple and inexpensive using either in vitro or in vivo hypermutation techniques~\cite{molina2022vivo}, the same is not true for obtaining genotype–phenotype linkages. The current state of the art relies on growth-based selection assays or fluorescence-activated cell sorting (FACS) coupled with next-generation sequencing to profile variants in the thousands to tens of thousands~\cite{fowler2014protocol, shin2023dms}. The exploration of this sequence space is also constrained to random mutations or restricted to combinations that are feasible to construct using contemporary molecular cloning techniques. This imposes a significant challenge in studying many evolutionary phenomena, such as, but not limited to, contingency, entrenchment, bifurcation, and neofunctionalization, during the evolution of extant enzymes within a protein family. Generative models, coupled with evolutionary simulations outlined in \autoref{chap:evodin}, can help alleviate this by (i) generating a representative library of the sequence space in silico and (ii) exploring how trajectories are navigated across a large subset of extant or ancestral proteins of interest. This enables researchers to empirically test a much smaller subset of sequences that are flagged by evolutionary simulations, rather than generating and screening massive sequence libraries. The experimental bottleneck then becomes gene synthesis, which has improved rapidly and continues to improve, in both cost and efficiency~\cite{plesa2020dropsynth2}.


One possible alternative application of these models concerns the construction of synthetic fitness landscapes spanning large protein families composed of multiple functional subfamilies. 
In this context, recent protein language models have demonstrated the ability to capture broad evolutionary and functional relationships across diverse sequence spaces, enabling the generation of proteins with novel or altered functions \cite{madani_large_2023, Nijkamp2022ProGen2ET, bhatnagar2025scaling}. Generative fitness-landscape models could provide a complementary framework by explicitly describing the fitness structure connecting different subfamilies and by identifying mutational trajectories that progressively shift function while preserving viable intermediate variants. 
With sufficiently accurate models, such approaches could guide enzyme optimization and functional re-engineering while reducing the need for exhaustive directed-evolution campaigns. 
A substantial effort is being made on steering sequence generation towards specific properties \cite{Calvanese2026steering}.




Finally, validating GSMs against evolving pathogens will be crucial for public health; such models could become essential assets in predicting dangerous variants and optimizing vaccine responses \cite{rodriguez2022epistatic, biswas2021low}. 
Beyond infectious diseases, GSMs may also find important applications in human genetics, where predicting the effects of genetic variants is a central challenge. 
In particular, accurate genotype-to-fitness mappings could help prioritize disease-associated mutations, identify cancer driver variants, and improve the interpretation of variants of uncertain significance. 
Recent community efforts, such as benchmarking initiatives based on large-scale variant-effect datasets, have highlighted the growing potential of generative models for these tasks~\cite{tabet2024benchmarking}. 
While current applications are often focused on the effects of individual mutations, extending these approaches to capture higher-order genetic interactions remains an important direction for future research.


\section*{Acknowledgments}
This study was supported by the Human Frontier Science Program (HFSP) Fellowship LT0052/2024-C (M.B.) (DOI: \url{https://doi.org/10.52044/HFSP.LT00522024-C.pc.gr.195454})
\newpage

\bibliography{references} 

\end{document}